\documentclass[lettersize,journal]{IEEEtran}
\usepackage{amsmath,amsfonts}
\usepackage{algorithmic}
\usepackage{algorithm}
\usepackage{array}
\usepackage[caption=false,font=normalsize,labelfont=sf,textfont=sf]{subfig}
\usepackage{textcomp}
\usepackage{stfloats}
\usepackage{url}
\usepackage{verbatim}
\usepackage{graphicx}
\usepackage{cite}
\usepackage{booktabs}       
\usepackage{multirow}
\begin{document}
	
	\title{CAMP-Net: Consistency-Aware Multi-Prior Network for Accelerated MRI Reconstruction}
	
	\author{Liping Zhang, Xiaobo Li, and Weitian Chen
		\thanks{This work was supported by a grant from Innovation and Technology Commission of the Hong Kong SAR (MRP/046/20X); and by a Faculty Innovation Award from the Faculty of Medicine of The Chinese University of Hong Kong. (Corresponding author: Weitian Chen.)}
		\thanks{Liping Zhang and Weitian Chen are with the Department of Imaging and Interventional Radiology, Faculty of Medicine, The Chinese University of Hong Kong, Hong Kong, China (e-mail: lpzhang@link.cuhk.edu.hk; wtchen@cuhk.edu.hk)}
		\thanks{Xiaobo Li is with School of Marine Science and Technology, Tianjin University, Tianjin, China (e-mail: lixiaobo@tju.edu.cn)}
	}
	
	\markboth{Journal of \LaTeX\ Class Files,~Vol.~X, No.~X, Dec~2023}%
	{Zhang \MakeLowercase{\textit{et al.}}: CAMP-Net: Consistency-Aware Multi-Prior Network for Accelerated MRI Reconstruction}
	
	
	\maketitle
	
	\begin{abstract}
		Undersampling $k$-space data in magnetic resonance imaging (MRI) reduces scan time but pose challenges in image reconstruction. Considerable progress has been made in reconstructing accelerated MRI. However, restoration of high-frequency image details in highly undersampled data remains challenging. To address this issue, we propose CAMP-Net, an unrolling-based Consistency-Aware Multi-Prior Network for accelerated MRI reconstruction. CAMP-Net leverages complementary multi-prior knowledge and multi-slice information from various domains to enhance reconstruction quality. Specifically, CAMP-Net comprises three interleaved modules for image enhancement, $k$-space restoration, and calibration consistency, respectively. These modules jointly learn priors from data in image domain, $k$-domain, and calibration region, respectively, in data-driven manner during each unrolled iteration. Notably, the encoded calibration prior knowledge extracted from auto-calibrating signals implicitly guides the learning of consistency-aware $k$-space correlation for reliable interpolation of missing $k$-space data. To maximize the benefits of image domain and $k$-domain prior knowledge, the reconstructions are aggregated in a frequency fusion module, exploiting their complementary properties to optimize the trade-off between artifact removal and fine detail preservation. Additionally, we incorporate a surface data fidelity layer during the learning of $k$-domain and calibration domain priors to prevent degradation of the reconstruction caused by padding-induced data imperfections. We evaluate the generalizability and robustness of our method on three large public datasets with varying acceleration factors and sampling patterns. The experimental results demonstrate that our method outperforms state-of-the-art approaches in terms of both reconstruction quality and $T_2$ mapping estimation, particularly in scenarios with high acceleration factors.
	\end{abstract}
	
	\begin{IEEEkeywords}
		Image restoration, image reconstruction, magnetic resonance imaging, biomedical image processing
	\end{IEEEkeywords}
	
	\section{Introduction}
	\label{sec:intro}
	\IEEEPARstart{M}{agnetic} resonance imaging (MRI), as a non-invasive and radiation-free medical imaging modality, is widely used in clinical and research settings. However, its slow acquisition can cause motion-related artifacts and patient discomfort, limiting its feasibility in certain applications. Accelerated MRI acquisitions by undersampling $k$-space offer reduced imaging scan times but often pose challenges in image reconstructions.
	
	Tremendous progress has been made to reconstruct accelerated MRI over the last few decades. Partial Fourier imaging \cite{Feinberg_1986} utilizes the conjugate symmetry property of $k$-space data for moderate acceleration. Parallel imaging (PI) techniques \cite{pruessmann1999sense, griswold2002generalized} leverage spatial information of coil sensitivities of phased array to speed up MRI acquisition. Compressed sensing MRI (CS-MRI) utilizes prior information such as sparsity \cite{lustig2007sparse} and low-rank \cite{dong2014compressive} properties of data in certain transform domains for image reconstruction. Additionally, low-rank structured matrix completion approaches like SAKE \cite{shin2014calibrationless}, LORAKS \cite{haldar2013low}, and ALOHA \cite{jin2016general} have been developed for fast MRI. 
	
	Recent advances in deep learning (DL) techniques have sparked significant interests in DL-based MRI reconstruction (DL-MRI). Many studies have leveraged image priors for artifact removal \cite{wang2016accelerating,yang2018admm,lee2018deep, wang2020deepcomplexmri}, while others have explored $k$-space correlations for missing data interpolation \cite{zhang2018multi,akccakaya2019scan,kim2019loraki,han2020k}. More recent research has focused on cross-domain methods that incorporate image prior knowledge with $k$-space information extracted from either traditional PI techniques \cite{ryu2021k,sriram2020grappanet} or dedicated neural networks \cite{eo2018kiki,wang2022dimension,ran2020md,sriram2020grappanet} for better reconstruction, achieving higher acceleration factors than using either domain knowledge alone. However, when incorporating $k$-space prior knowledge into DL-MRI, these methods may face limitations of restoration of high-frequency details, particularly at highly accelerated MRI. Firstly, a limitation arises from capturing the inherent correlations in multi-coil $k$-space data. Methods have been developed to use networks to capture $k$-space correlations by data consistency layers to ensure data fidelity of acquired data \cite{schlemper2017deep,zhang2018multi}, but they lack sufficient constraints to ensure accurate estimation of missing data. Scan specific methods attempt to utilize auto-calibration signal (ACS) data \cite{akccakaya2019scan,kim2019loraki}, but these methods may face challenges of limited calibration data and the presence of noise, especially at high acceleration factors. Furthermore, fixed local kernels learned from ACS data offline, rather than kernels optimized jointly with the network during the learning process, are often used \cite{sriram2020grappanet,ryu2021k,ryu2022improving}, resulting in marginal improvement. Secondly, the utilization of inter-slice correlations in the $k$-domain is not well explored. Similar to inter-slice correlations in the image domain \cite{song2014reconstruction,shangguan2022multi,sun2019deep,fabian2022humus}, utilizing correlation information between the target slice and its adjacent slices in the $k$-domain can benefit reconstruction. However, existing approaches \cite{du2020multiple} that utilize Convolutional Neural Networks (CNNs) to capture local dependencies from adjacent $k$-space slices face challenges in handling artificial signals around padding areas. These imperfections can affect $k$-space signal synthesis and disrupt inter-slice correlations, resulting in data misinterpretation and inaccurate restoration of subtle features.
	
	In this study, we propose a novel unrolling-based Consistency-Aware Multi-Prior Network (CAMP-Net) for reconstruction of accelerated MRI. CAMP-Net effectively addresses the challenges posed by existing methods by integrating PI-MRI and CS-MRI techniques within a deep learning framework to jointly utilize priors in image domain, $k$-domain, and calibration regions. By collaboratively learning these diverse priors in a data-driven fashion, CAMP-Net captures their complementary properties for robust MRI reconstruction, even in the presence of data imperfections. One advantage of CAMP-Net is its incorporation of data-adaptive self-consistent calibration priors, which guides the modeling of consistency-aware $k$-space correlations and ensure consistent interpolation of missing data. Notably, we maximize the utilization of calibration information by leveraging the end-to-end learning capability of networks, distinguishing it from conventional offline approaches \cite{sriram2020grappanet,ryu2021k,ryu2022improving}. Moreover, CAMP-Net employs a frequency fusion layer that combines the outputs of image and $k$-space priors in the frequency domain. This enables error back-propagation to the specific prior, facilitating the learning of their distinct properties and achieving an improved trade-off between artifact removal and fine detail preservation. Furthermore, CAMP-Net sequentially explores adjacent slice redundancy across image, $k$-space, calibration domains, enhancing the learning of image, $k$-space, and calibration priors by capturing intrinsic intra- and inter-slice features. In particular, a surface data fidelity layer is proposed to prevent back-propagated errors of imperfect data from affecting the learning of $k$-space and calibration priors during the exploration of adjacent slice information in their frequency domains.
	In summary, the key contributions of this work are:
	\begin{itemize}
		\item We propose CAMP-Net for the reconstruction of accelerated MRI by leveraging complementary multi-prior knowledge from images, $k$-space data, and calibration region to improve the quality of reconstructed images.
		\item We establish reliable and consistency-aware $k$-space correlations by incorporating both image and calibration prior knowledge through a cohesive feature learning.
		\item A frequency fusion layer is used to aggregate information from image and $k$-space priors, optimizing the utilization of their distinct properties to improve reconstruction.
		\item We exploit the multi-slice data redundancy across image, $k$-space, calibration domains to enhance multi-prior knowledge. Specifically, we incorporate a surface data fidelity layer to mitigate the adverse impacts of noisy data on learning $k$-space and calibration priors.
		\item Extensive experiments on three public datasets under various acceleration factors demonstrate that the proposed CAMP-Net outperforms the existing state-of-the-art methods in terms of tissue structure restoration, aliasing artifact removal, and $T_2$ parameter estimation.
	\end{itemize}
	
	\section{Methods}
	\label{sec:methods}
	We first formulate Parallel MRI reconstruction as an inverse problem, solvable through our learnable multi-prior optimization formulation. Then, we introduce CAMP-Net and elaborate on its core modules.
	
	\subsection{Problem Formulation for Parallel MRI Reconstruction}
	Parallel MRI uses radiofrequency (RF) coils to simultaneously encode spatial information of an image $x \in \mathbb{R}^N$. Noise-free measurements $\tilde{y} \in \mathbb{C}^{N_cN}$ are obtained by sampling signals of $N_c$ coils in the frequency domain for image reconstruction, which can be mathematically expressed as:
	\begin{equation}
		\tilde{y} = \mathcal{MFS}x = \mathcal{A}x,
		\label{eq:forward_process}
	\end{equation}
	where $\mathcal{A}: \mathbb{R}^N \to \mathbb{C}^{N_cN}$ is the forward process that involves sequential operators of coil sensitivity map projection $\mathcal{S}: \mathbb{R}^N \to \mathbb{C}^{N_cN}$, Fourier transform $\mathcal{F}: \mathbb{C}^{N_cN} \to \mathbb{C}^{N_cN}$, and under-sampling pattern $\mathcal{M}: \mathbb{C}^{N_cN} \to \mathbb{C}^{N_cN}$. Coil sensitivity maps can be acquired or estimated from fully-sampled auto-calibration signal (ACS) data acquired during the same scan.
	
	\begin{figure}[!t]
		\centering
		\includegraphics[width=0.8\columnwidth]{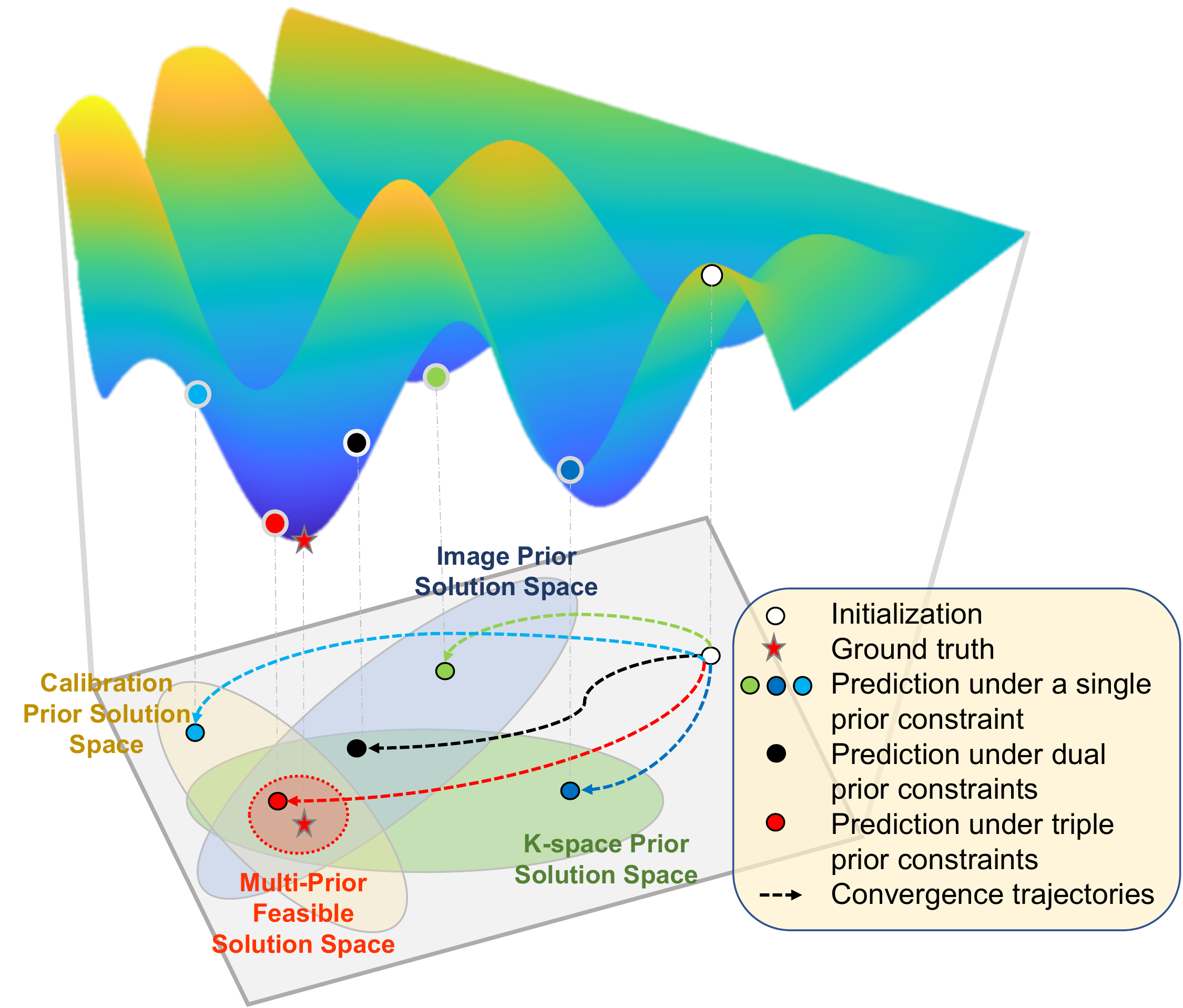}
		\caption{
			Convergence trajectories of reconstruction under different conditions.
		}
		\label{fig:convergence-trajectory}
	\end{figure}
	
	Reconstructing the desired image from under-sampled $k$-space data is an ill-posed inverse problem. Conventional CS-MRI methods \cite{lustig2008compressed} leverage predefined prior knowledge to regularize the inverse problem. Instead, we approach it as a learnable multi-prior optimization problem, formulated as:
	\begin{equation}
		\hat{x} = \arg\min_{x} \| \mathcal{A}x - \tilde{y} \| ^2_2 + \sum_{l=1}^{L}\lambda_l \mathcal{P}(\mathcal{D}_{l}x;\theta_l),
		\label{eq:multi_prior_optimization}
	\end{equation}
	where $\mathcal{P}(\mathcal{D}_{l}x;\theta_l)$ is a data-adaptive prior with learnable parameters $\theta_l$ that incorporate learned knowledge in the specific transform domain $\mathcal{D}_{l}$, and $\lambda_l$ balances the influence of the imposed prior and the fidelity of the acquired data. The advantages of utilizing multiple learnable prior constraints for reconstruction are illustrated in Fig.~\ref{fig:convergence-trajectory}. The presence of numerous feasible sets in the solution space poses a challenge for searching the optimal reconstruction with a single prior. This often leads to convergence towards local minima that may satisfy one constraint but violate another, making them sensitive to noise and initialization conditions. In contrast, the interaction of multiple priors provides collaborative adjustments toward convergence during training, enabling a more comprehensive and balanced optimization process. This approach effectively leverages the complementary properties of the priors, leading to stable convergence and a more reliable solution space for the inverse problem.
	
	\subsection{Overall Architecture of CAMP-Net}
	\begin{figure*}[!t]
		\centering
		\includegraphics[width=\textwidth]{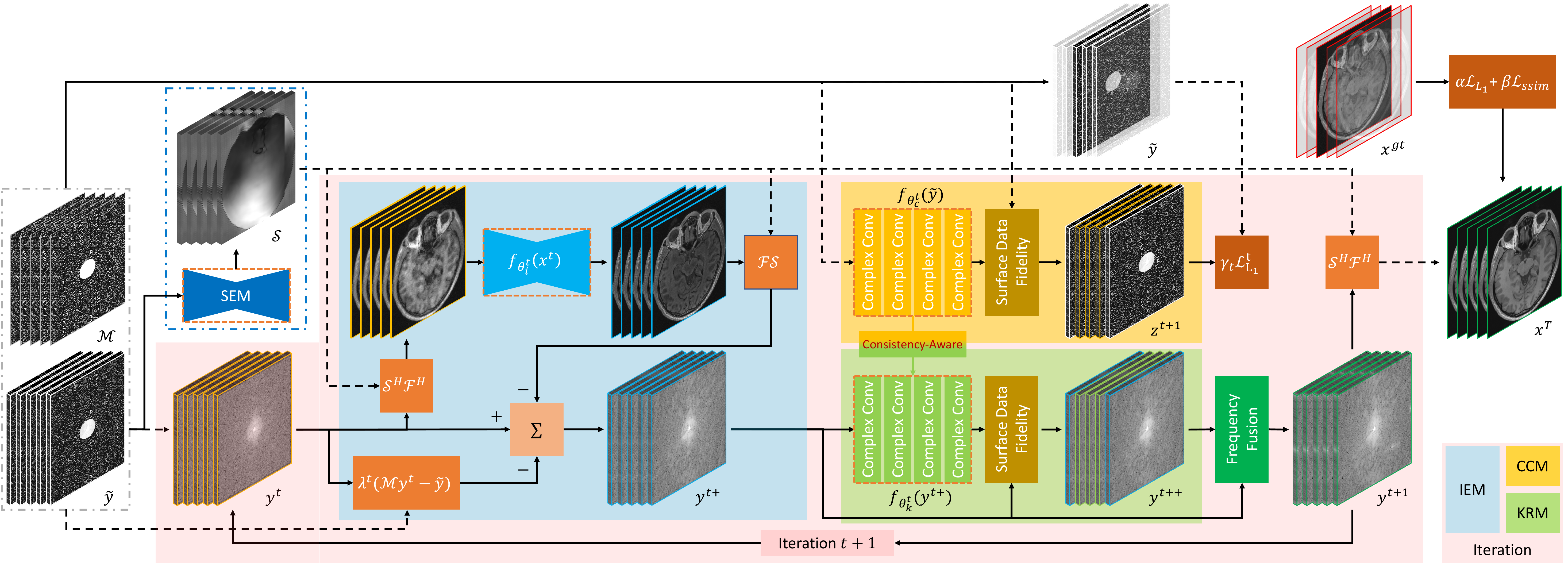}
		\caption{
			Architecture of CAMP-Net for accelerated MRI reconstruction. CAMP-Net takes as input under-sampled k-space data and sampling masks of consecutive slices. With coil sensitivity maps from the SEM module, the IEM, KRM, CCM, and FFM modules iteratively reconstruct the slices. Only the center slice undergoes direct supervision via its reconstruction loss; adjacent slices of the reference appear transparent, indicating indirect supervision. Solid and dashed lines represent information flow, with dashed lines depicting inputs external to iterations interacting internally within the modules.
		}
		\label{fig:overall-architecture}
	\end{figure*}
	
	CAMP-Net is an unrolling-based approach for solving \eqref{eq:multi_prior_optimization}, where image, $k$-space, and calibration prior knowledge are mutually constrained and learned in each iteration to achieve progressive reconstruction. The architecture of CAMP-Net, illustrated in Fig.~\ref{fig:overall-architecture}, consists of several key modules: Spatial Encoding (SEM), Image Enhancement (IEM), $k$-space Restoration (KRM), Calibration Consistency (CCM), and Frequency Fusion (FFM). Additionally, CAMP-Net effectively captures both intra- and inter-slice correlations within each prior by incorporating the target slice and its adjacent slices as input during each iteration.
	
	\begin{algorithm}[H]
		\caption{CAMP-Net}
		\label{algorithm:1}
		\begin{algorithmic}
			\STATE
			\renewcommand{\algorithmicrequire}{\textbf{Input:}}
			\renewcommand{\algorithmicensure}{\textbf{Output:}}
			\REQUIRE $\tilde{y}$ the multi-coil $k$-space measurements,
			$\mathcal{M}$ the binary sampling pattern, $\mathcal{M}_k$ the surface mask for KRM, $\mathcal{M}_c$ the surface mask for CCM, and $T$ the unroll iteration.
			\ENSURE	$x$ the reconstructed image
			\STATE	$y^0 = \tilde{y}$; // initialization
			\STATE	$\mathcal{S} = \text{SEM}(y^0;\theta_s^t)$; // sensitivity maps estimation
			\FOR {$t = 0$ to $T-1$}
			\STATE $y^{t+} = \text{IEM}(y^{t}, \tilde{y}, \mathcal{M}, \mathcal{S};\theta_i^t)$; // image enhancement
			\STATE	$y^{t++} = \text{KRM}(y^{t+}, \mathcal{M}_k; \theta_k^t)$; // $k$-space restoration
			\STATE	$z^{t+1} = \text{CCM}(\tilde{y}, \mathcal{M}_c; \theta_c^t)$; // calibration consistency
			\STATE $\theta_c^t \gets \to \theta_k^t$; // weight sharing mechanism
			\STATE	$y^{t+1} = \text{FFM}(y^{t+}, y^{t++}, \mathcal{M})$; // frequency fusion
			\ENDFOR
			\RETURN $x = \mathcal{S}^H\mathcal{F}^Hy^{T}$; // final reconstructed image
		\end{algorithmic}
	\end{algorithm}
	
	The pseudo algorithm is presented in Algorithm \ref{algorithm:1}. Specifically, the SEM is used to learn coil sensitivity maps from the ACS data. These sensitivity maps are then utilized by CAMP-Net during each iterative learning process. To initiate each iterative learning process, CAMP-Net utilizes the IEM to extract image prior knowledge from coil-combined images. By prioritizing the learning process of image priors compared to $k$-space priors in CAMP-Net, we can mitigate the adverse impact from missing data on learning $k$-space correlations~\cite{souza2019hybrid}. Building upon the outputs of the image prior learning process, the KRM module in CAMP-Net utilizes information from multi-coil complex images and $k$-space correlations to obtain missing data from neighboring $k$-space data. Additionally, CAMP-Net leverages the CCM to embed calibration information from ACS data and propagates consistent representations to enforce KRM to learn trustworthy neighborhood relationships and vice versa. Moreover, at the end of each iteration, the information flows of IEM and KRM interact via FFM, combining their mutual advantages for improved reconstruction. Finally, CAMP-Net generates the final reconstructed image by applying coil combination techniques, such as the Root Sum of Squares (RSS) or sensitivity-like projection, to the output $k$-space data of the final iteration.
	
	\subsection{Contextual Information from Adjacent Slices}
	Exploring data redundancy between the target slice and its adjacent slices benefits image reconstruction \cite{song2014reconstruction, fabian2022humus} as nearby images often has continuous anatomical structures. The continuity of image content in adjacent slices implies that adjacent $k$-space data often share similar information, leading to improved reconstruction \cite{du2020multiple}. Given a volume of under-sampled multi-coil $k$-space data $\tilde{y}_\text{vol}=(\tilde{y}_1,\cdots,\tilde{y}_{ns})$ with a matrix dimension of ($ns$, $nc$, ${k}_{y}$, ${k}_{x}$), the corresponding coil-combined image volume is $\tilde{x}_\text{vol}=(\tilde{x}_1,\cdots,\tilde{x}_{ns})$ with a matrix dimension of ($ns$, ${i}_{y}$, ${i}_{x}$). Here, $ns$ and $nc$ represent the number of slices and coils, and ${k}_{y}$, ${k}_{x}$, ${i}_{y}$, ${i}_{x}$ denote the matrix size in the y-axis ($k$-space), x-axis ($k$-space), y-axis (image), and x-axis (image), respectively. The target slice $\tilde{y}_{s}$ and its m-adjacent slices (AS) on each side can be expressed as $(\tilde{y}_{s-m},\cdots,\tilde{y}_{s-1},\tilde{y}_{s},\tilde{y}_{s+1},\cdots,\tilde{y}_{s+m})$, and the corresponding coil-combined adjacent image slices is $(\tilde{x}_{s-m},\cdots,\tilde{x}_{s-1},\tilde{x}_{s},\tilde{x}_{s+1},\cdots,\tilde{x}_{s+m})$. CAMP-Net utilizes 3D convolution kernels to capture contextual information from $2m+1$ consecutive slices. In CAMP-Net, multi-slice data are organized by concatenating slices along the depth dimension, while data from multiple receivers are stacked along the channel dimension for convolution. This enables CAMP-Net to learn both intra-slice and inter-slice features, capturing prior knowledge within and across slices. Notably, only the center slice is supervised during training, while its adjacent slices can leverage the learned priors without explicit supervision. 
	
	\subsection{Learning Image Enhancement}
	The IEM explores the underlying properties of coil-combined images by learning a data-driven image prior. It aims to restore the overall anatomical structure from degraded data by imposing desired image properties learned from the training data, generating artifact-free images and providing complete $k$-space signals for the subsequent $k$-space restoration. The module adopts an unrolled gradient descent (GD) algorithm that ensures data consistency with the acquired $k$-space samples while imposing the learned image prior on coil-combined images. The update formula at the $t$-th iteration is:
	
	\begin{equation}
		\begin{aligned}
			y^{t+} &= \text{IEM}(y^{t}, \tilde{y}, \mathcal{M}, \mathcal{S};\theta_i^t) \\
			&= y^{t} - \eta^{t} \big(\mathcal{FS}f_{\theta_i^t}(\mathcal{S}^H\mathcal{F}^Hy^t) + \lambda^t (\mathcal{M}y^{t} - \tilde{y}) \big),
			\label{eq:pi_image_domain_iteration}
		\end{aligned}
	\end{equation}
	where $\theta_i^t$, $y^{t}$, $\eta^{t}$, and $\lambda^t$ denote the network parameters of $f_{\theta_i^t}$, initial multi-coil $k$-space data, learning rate, learnable regularization parameter at the $t$-th iteration, respectively. The network architecture for $f_{\theta_i^t}$ in the IEM extends the 2D U-Net from \cite{sriram2020end} to 3D convolutions. This enables modeling inter-slice correlations and multi-slice context. Similarly, the SEM utilizes this 3D network to estimate slice-dependent sensitivity maps $\mathcal{S}$ by capturing inter-slice coil profile dependencies.

	\subsection{Learning $k$-space Restoration}
	\begin{figure*}[!t]
		\centering
		\includegraphics[width=0.96\textwidth]{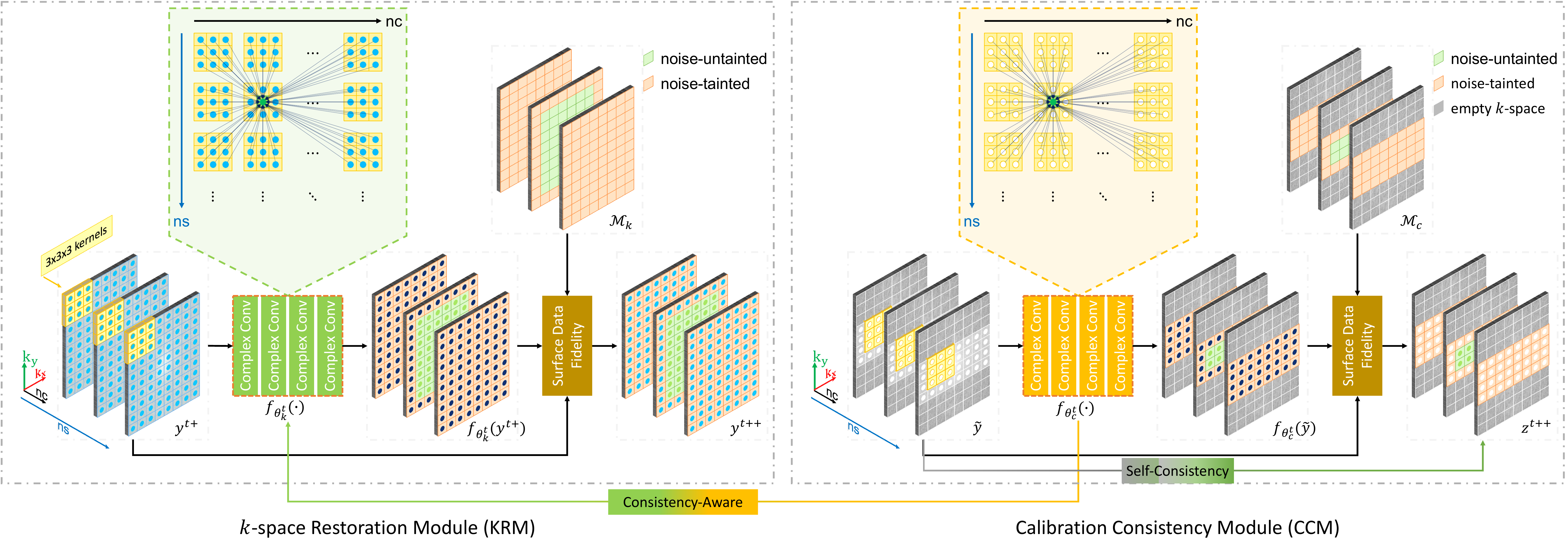}
		\caption{
			Details of KRM and CCM. The KRM takes multi-slice multi-coil $k$-space data as input and utilizes a complex-valued network to refine the data, along with a surface data fidelity layer to handle data imperfections. The CCM takes multi-slice multi-coil ACS data as input and utilizes another complex-valued network and surface data fidelity layer to encode calibration information from the input data to guide the KRM in learning consistency-aware correlations.
		}
		\label{fig:krm-ccm-flowchart}
	\end{figure*}
	
	The IEM excels in removing noise and aliasing artifacts from degraded images, but its reconstructions may lack fine-scale texture and structural details due to over-smoothing. Preserving high-frequency information is crucial for many tasks of medical diagnosis. Previous works exploring $k$-space correlations for missing signal recovery have advantages in maintaining such details and improving overall image quality \cite{ryu2021k, sriram2020grappanet}. However, utilizing fixed and pre-defined kernels in these approaches has limited their capacity to represent complex features and generalize to varying data distributions. To address such limitations, we propose a novel module called KRM that learns a data-driven $k$-space prior optimized jointly with the network during training. It models underlying frequency spectra relationships to refine IEM outputs, capturing high-frequency structural details and fine-scale textures. The KRM comprises a complex-valued network for modeling $k$-space correlations and a surface data fidelity (SDF) layer for reducing data imperfections. The details of the KRM are shown in Fig.~\ref{fig:krm-ccm-flowchart}. Its update formula in the $t$-th iteration is:
	\begin{equation}
		y^{t++}
		= \text{KRM}(y^{t+}, \mathcal{M}_k;\theta_k^t)
		= \text{SDF}(f_{\theta_k^t}(y^{t+}), y^{t+}, \mathcal{M}_k),
	\end{equation}
	where $\theta_k^t$ is the network parameters of the $f_{\theta_k^t}$ at the current iteration and $\mathcal{M}_k$ is a surface mask for the KRM.
	
	\subsubsection{Learning $k$-space Correlations}
	The potential benefits of using complex-valued networks for multi-channel image reconstruction have been demonstrated in \cite{wang2020deepcomplexmri}. In this study, we design a specialized complex-valued network to capture $k$-space correlations from a set of consecutive slices of multi-coil $k$-space data, with the aim of fully exploiting the information contained in the complex-valued multi-slice and multi-coil data. As illustrated in Fig.~\ref{fig:krm-ccm-flowchart}, the network comprises four complex-valued 3D convolution layers that use 3 $\times$ 3 $\times$ 3 kernels to extract features from $k$-space data, where the number of coils ($nc$) and slices ($ns$) are considered as the feature and depth dimensions, respectively. The data of each coil from each slice is estimated by utilizing information from all coils across all slices. This enables the network to capture complex intra- and inter-slice multi-coil features for high-frequency information restoration.
	
	\subsubsection{Surface Data Fidelity Layer}
	Padding operations are employed in the $f_{\theta_k^t}$ network to preserve spatial resolution by adding extra $k$-space values around the surfaces of the input $k$-space data during convolutions. However, the inclusion of these padding values introduces artifacts within the surface regions of the outputs, which can potentially hinder the learning of $k$-space correlations. This can lead to flawed frequency interpolation that spreads throughout the entire image, resulting in undesired reconstruction outcomes. This issue is particularly problematic in multi-slice scenarios, where padding along slices can introduce numerous imperfect data pairs that compromise or even disrupt the correlations between adjacent $k$-space slices.
	To reduce the influence of padding-induced artificial signals during $k$-space correlation learning, we propose a surface data fidelity (SDF) layer, as illustrated in Fig.~\ref{fig:krm-ccm-flowchart}.
	This layer enforces hard data fidelity with the input of the KRM, denoted as $y^{t+}$, specifically targeting the imperfect surfaces while preserving the noise-untainted data in the refined results, denoted as $f_{\theta_k^t}(y^{t+})$. The general formula for the SDF is as follows:
	\begin{equation}
		\text{SDF}(f_{\theta_k^t}(y^{t+}), y^{t+}, \mathcal{M}_{k})
		= \mathcal{M}_{k}f_{\theta_k^t}(y^{t+}) + (1-\mathcal{M}_{k})y^{t+},
	\end{equation}
	where $\mathcal{M}_{k}$ is a binary surface mask for the KRM that preserves noise-untainted areas of the input $f_{\theta_k^t}(y^{t+})$ while enforcing data fidelity on the noise-tainted areas of the input $y^{t+}$. Note that the noise-tainted areas include the intra-slice boundaries affected by padding operations, as well as the outer two slices in multi-slice scenarios, as shown in Fig.~\ref{fig:krm-ccm-flowchart}.
	
	\subsection{Learning Scan-Specific Calibration Consistency}
	Shift-invariant $k$-space correlations derived from ACS data have been extensively investigated in conventional PI techniques \cite{griswold2002generalized}. Recent DL-based MRI methods leverage this information for scan-specific reconstruction \cite{akccakaya2019scan,kim2019loraki} or as additional constraints to improve deep learning model outputs \cite{ryu2021k,sriram2020grappanet}. In contrast, we propose a novel approach using the CCM to encode calibration information and implicitly guide the KRM in learning consistency-aware $k$-space correlations. The details are shown in Fig.~\ref{fig:krm-ccm-flowchart}. The CCM and KRM share the same architecture, but the CCM takes a series of consecutive slices of the multi-coil ACS data as input and predicts the input itself by enforcing a self-consistency constraint. To effectively incorporate the calibration prior learned from the CCM, we establish a consistency-aware mechanism by sharing network parameters between the CCM and the KRM during end-to-end learning. This enables the KRM to simultaneously integrate the calibration prior and the image-enhanced intermediate information, resulting in robust $k$-space correlations. As a result, our approach enables consistent prediction of both acquired and missing $k$-space data . Furthermore, to mitigate the influence of padding-induced artificial signals that can lead to inaccurate calibration prior, we incorporate the SDF layer in the CCM. This layer selectively focuses on the inner calibration regions of the ACS data volume while excluding its surface regions during network back-propagation.
	The update procedure of the CCM at the $t$-th iteration is as follows:
	\begin{equation}
		z^{t+1}
		= \text{CCM}(\tilde{y}, \mathcal{M}_{c};\theta_c^t)
		= \mathcal{M}_{c}f_{\theta_c^t}(\tilde{y}) + (1-\mathcal{M}_{c})\tilde{y},
	\end{equation}
	where $\theta_c^t$ is the network parameters of $f_{\theta_c^t}$ at the $t$-th iteration and $\mathcal{M}_c$ is a surface mask for the CCM.
	
	\subsection{Frequency Fusion Module}
	The IEM and KRM are designed to excel in restoring low-frequency features for generating artifact-free images and preserving high-frequency signals in the frequency domain, respectively. To leverage the benefits of both modules, we propose the FFM to aggregate their predictions in the frequency domain. This module enables the network to optimize and leverage the distinct properties of various priors to improve reconstruction. The process at the $t$-th iteration is:
	\begin{equation}
		y^{t+1}
		= \text{FFM}(y^{t+}, y^{t++}, \mathcal{M})
		= \mathcal{M}y^{t+} + (1-\mathcal{M})y^{t++},
	\end{equation}
	where $\mathcal{M}$ is the sampling mask and $y^{t+1}$ denotes the final output of the $t$-th iteration, which is selected from the image-enhanced data $y^{t+}$ and the $k$-space refined data $y^{t++}$. Consequently, the forward information from the two modules is interactively merged in every iteration to ensure that the reconstruction has fine-scale textures and structural details.
	
	\subsection{Learning and Evaluation Strategy}
	\subsubsection{Loss Functions}
	To minimize the difference between the reconstructed images and the ground truth, we employ $L_1$ and Structural Similarity Index Measure (SSIM) loss functions. Additionally, a $L_1$ calibration loss is used to enforce the CCM to embed self-consistent features in each iteration.
	Note that supervision is provided only to the center slice of adjacent slices. The total losses for optimization are as follows:
	\begin{equation}
		\mathcal{L} = \alpha\mathcal{L}_{L_1}(x^T, x^{gt}) + \beta\mathcal{L}_\text{SSIM}(x^T,x^{gt}) + \sum_{t=1}^T \gamma_t \mathcal{L}^{t}_{L_1}(z^{t},\tilde{y}),
	\end{equation}
	where $x^T$ and $x^{gt}$ are the RSS or sensitivity-like reconstruction and the reference image, respectively. $z^{t}$ represents the output of the $(t-1)$-th CCM. The trade-off parameters $\alpha$, $\beta$, and $\gamma_t$ ($1 \leq t \leq T$) are all set to 1 during training for simplicity.
	
	\subsubsection{Evaluation Metrics}
	We evaluate reconstruction quality using the metrics of Peak Signal-to-Noise Ratio (PSNR) and SSIM. Additionally, the SKM-TEA dataset \cite{desai2022skm} includes MRI data for $T_2$ mapping, enabling us to compare the accuracy of $T_2$ quantification from the reconstructed MRI images.
	
	\section{Experiments and Results}
	\label{sec:experiments_and_results}
	\subsection{Datasets}
	\begin{figure}[!t]
		\centering
		\subfloat[]{
			\includegraphics[height=0.2\linewidth]{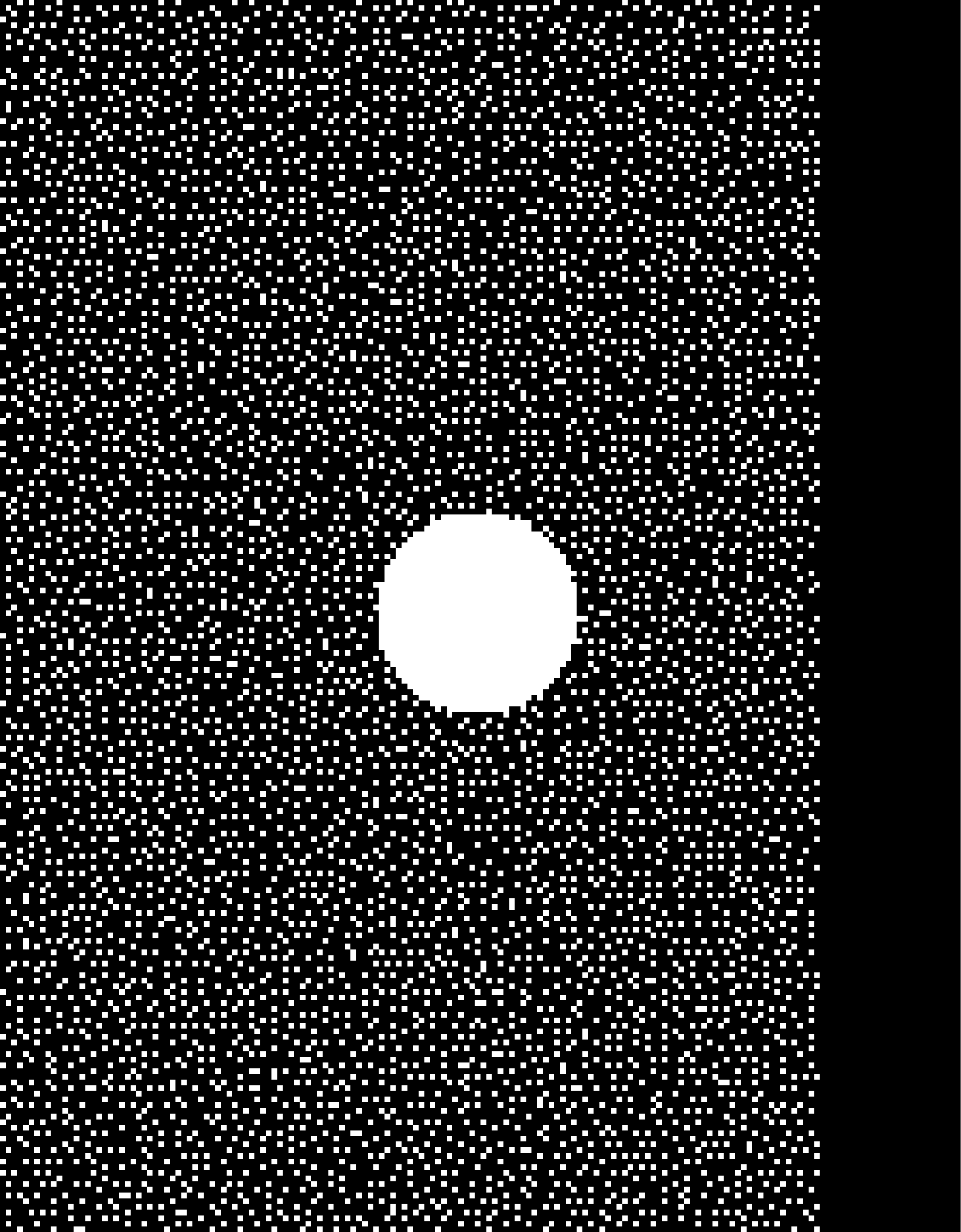}
			\label{fig:sampling_mask_a}
		}
		\subfloat[]{
			\includegraphics[height=0.2\linewidth]{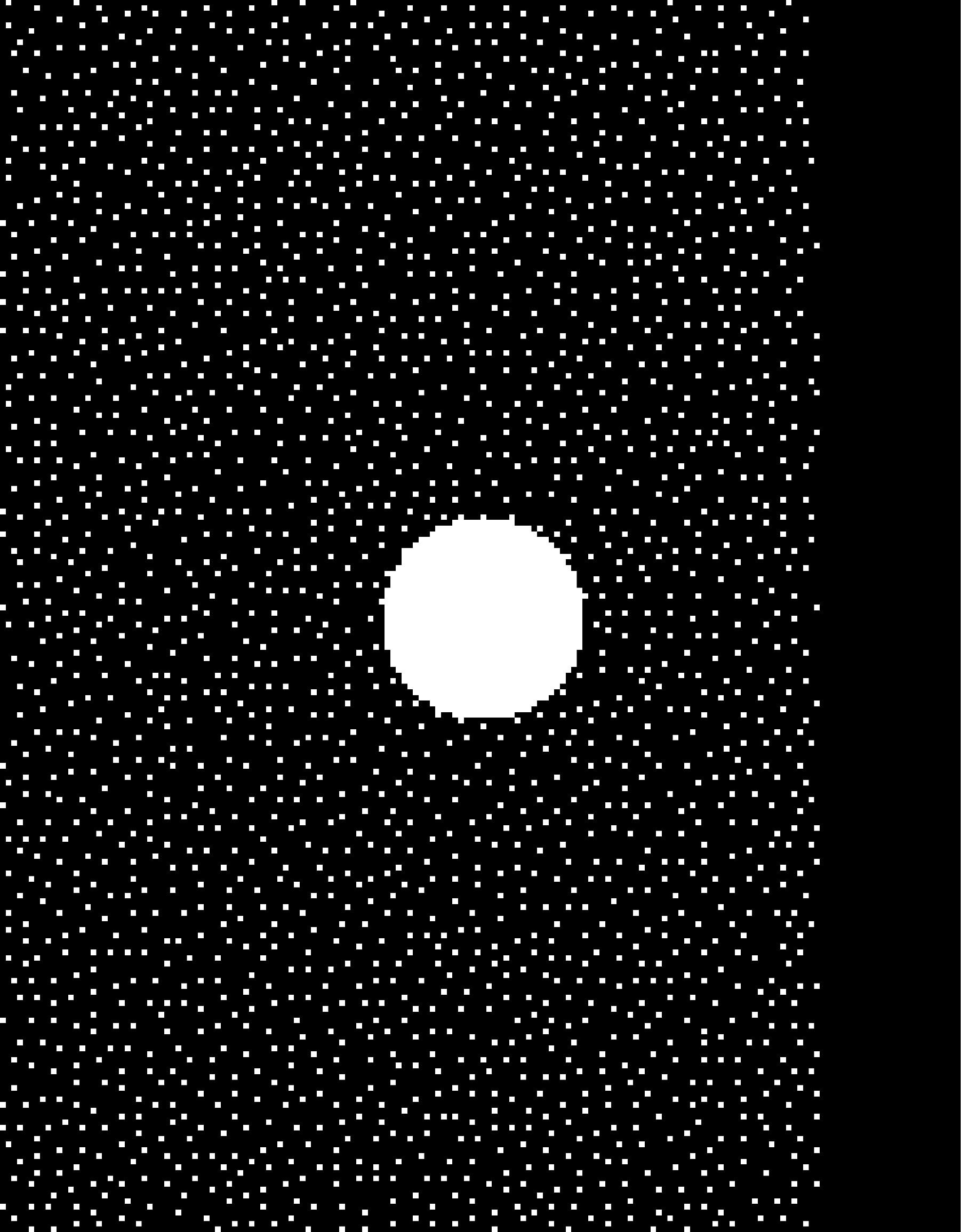}
			\label{fig:sampling_mask_b}
		}
		\subfloat[]{
			\includegraphics[height=0.2\linewidth]{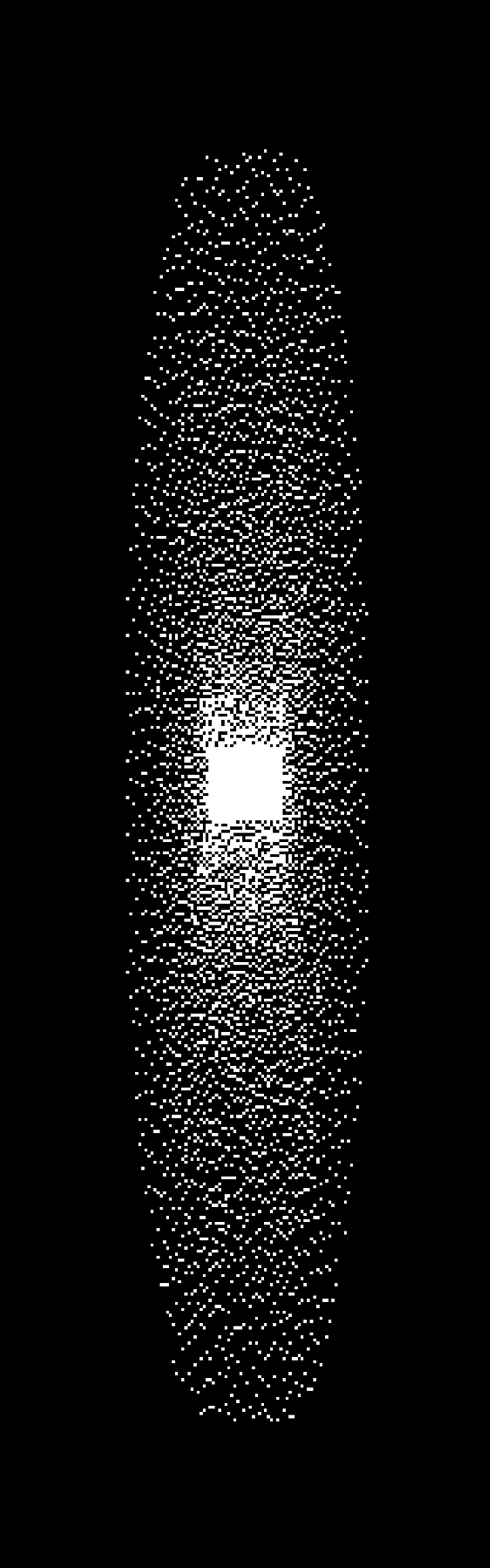}
			\label{fig:sampling_mask_c}
		}
		\subfloat[]{
			\includegraphics[height=0.2\linewidth]{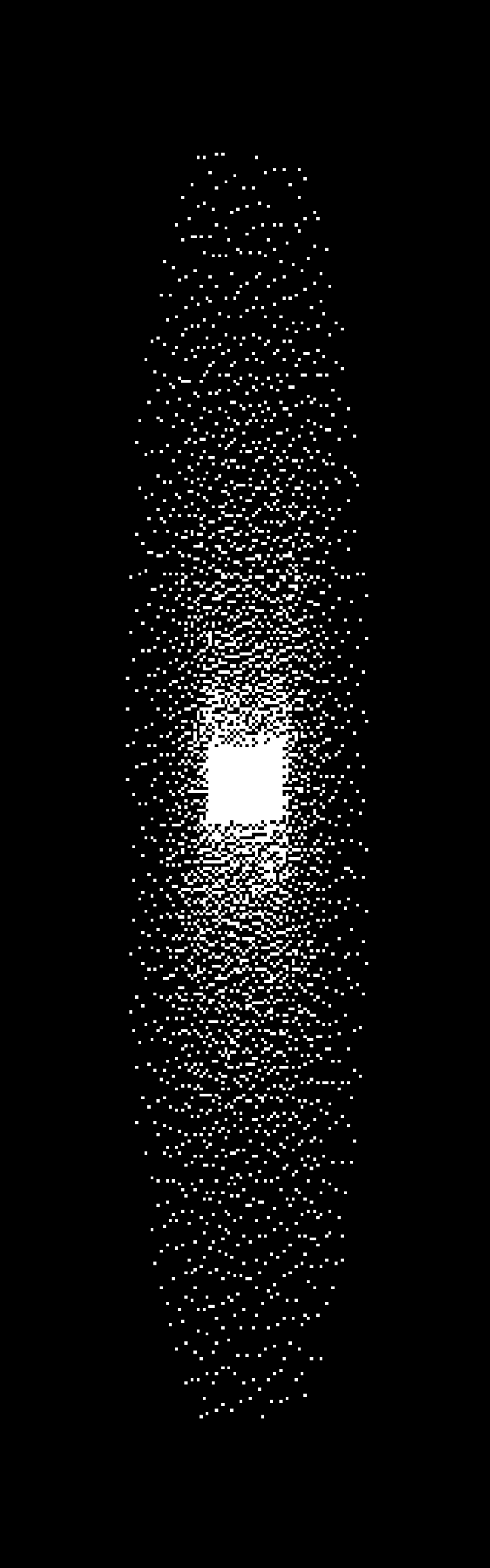}
			\label{fig:sampling_mask_d}
		}
		\subfloat[]{
			\includegraphics[height=0.2\linewidth]{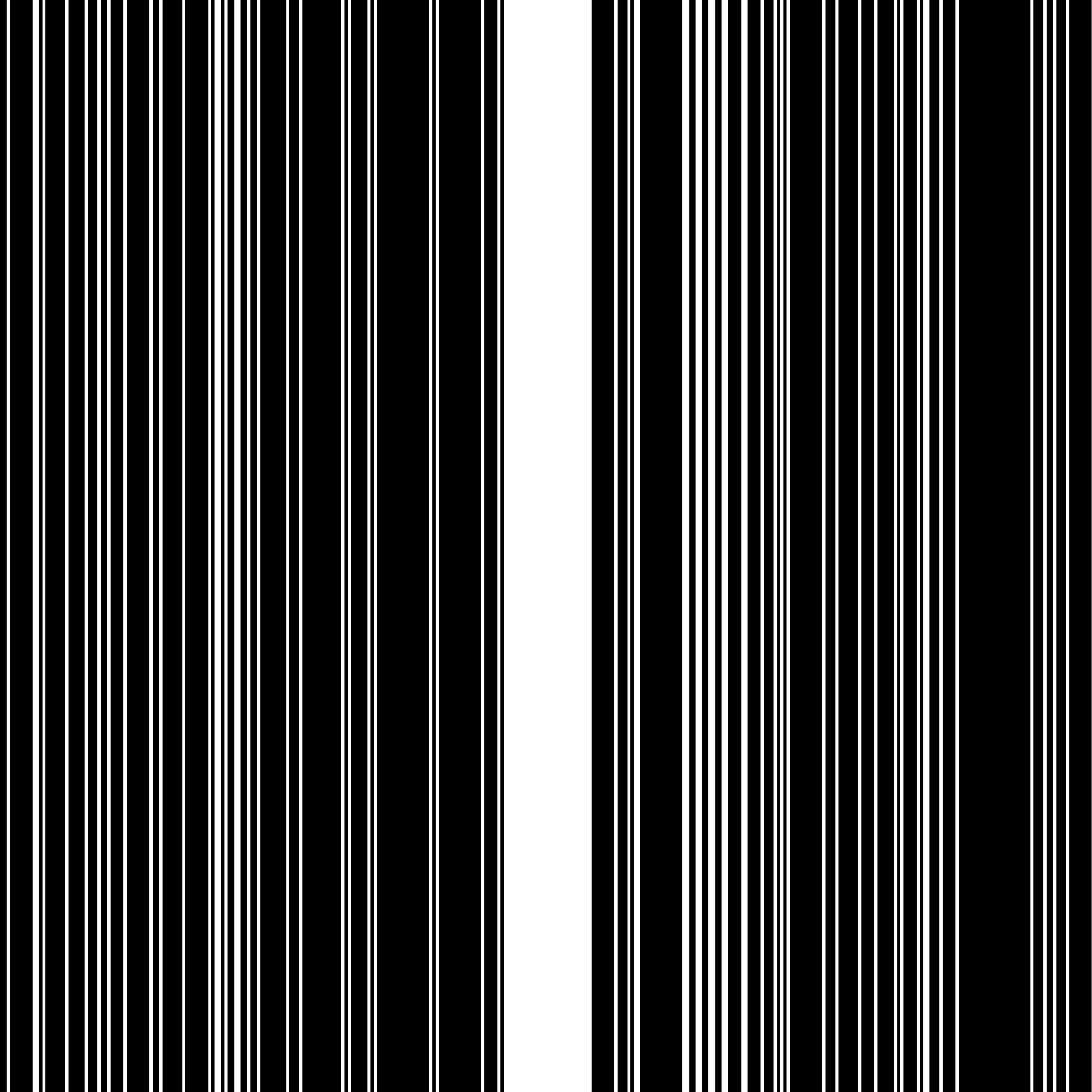}
			\label{fig:sampling_mask_e}
		}
		\subfloat[]{
			\includegraphics[height=0.2\linewidth]{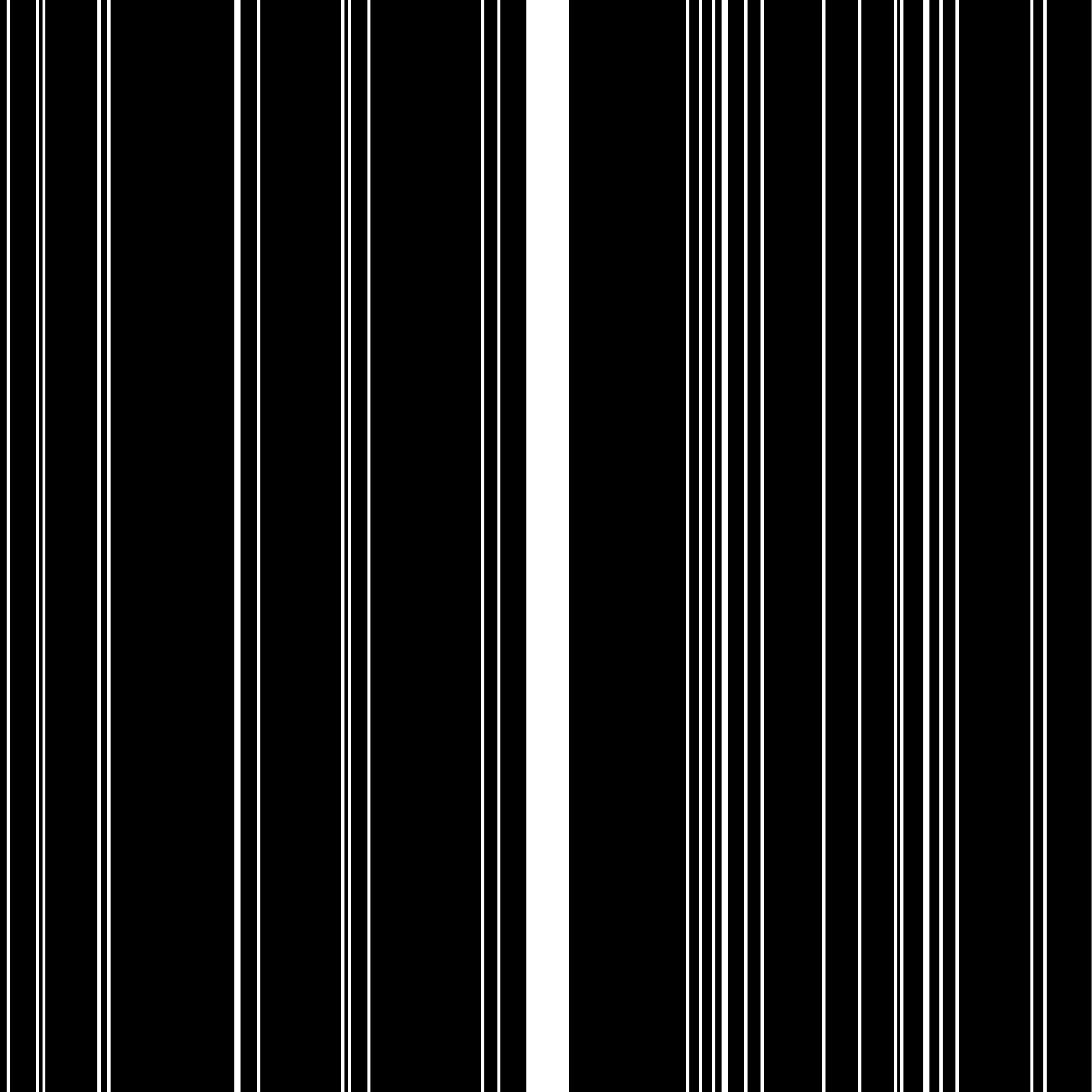}
			\label{fig:sampling_mask_f}
		}
		\caption{
			Sampling patterns: (a)-(b) 2D Poisson Disc sampling mask at acceleration factors of 5X and 10X for the Calgary-Campinas dataset, with an ACS region within a radius of 16. (c)-(d) 2D Poisson Disc undersampling at acceleration factors of 6X and 8X for the SKM-TEA dataset, with a kernel width of 6 and an ACS region of size 24$\times$24. (e)-(f) 1D random sampling mask at acceleration factors of 4X and 8X for the fastMRI Knee dataset, with ACS lines covering 8\% and 4\% of the total $k$-space phase lines, respectively.
		}
		\label{fig:sampling_mask}
	\end{figure}
	
	\subsubsection{Calgary-Campinas}
	We used the multi-coil brain MR raw data from the Calgary-Campinas dataset \cite{souza2018open}, which was collected using T1-weighted 3D gradient-recalled echo acquisitions with 1~mm isotropic resolution. The experiment was conducted on the provided split sets \cite{yiasemis2022recurrent}, including 47 volumes (7,332 axial slices) for training, 10 volumes (1,560 axial slices) for validation, and 10 volumes (1,560 axial slices) for testing. Each volume consists of fully-sampled $k$-space data acquired using a 12-channel coil. The data were partially collected up to 85\% of its matrix size in the slice-encoding direction (kz). We evaluated acceleration factors of 5 and 10 using Poisson disk distribution sub-sampling masks \cite{beauferris2022multi}. Examples of the masks are shown in Fig.~\ref{fig:sampling_mask}~\subref{fig:sampling_mask_a} and Fig.~\ref{fig:sampling_mask}~\subref{fig:sampling_mask_b}.
	
	\subsubsection{SKM-TEA}
	The SKM-TEA dataset consists of knee MRI scans from 155 patients who underwent the 3D qDESS scan on 3T GE MR750 scanners \cite{desai2022skm}. The test set includes data from 36 patients who received arthroscopic surgical intervention. The remaining data were randomly split into 86 scans for training and 33 scans for validation. Raw $k$-space data were acquired in a multi-coil setting (8 or 16 coils) with 2 $\times$ 1 PI using elliptical sampling. The fully-sampled $k$-space data was synthesized using ARC from the under-sampled measurements. Acceleration factors of 6 and 8 were investigated using a 2D Poisson Disc undersampling mask. Examples of the masks are shown Fig.~\ref{fig:sampling_mask}~\subref{fig:sampling_mask_c} and Fig.~\ref{fig:sampling_mask}~\subref{fig:sampling_mask_d}.
	
	\subsubsection{fastMRI Knee}
	We utilized the multi-coil knee MR raw data from the fastMRI competition \cite{zbontar2018fastmri}. The dataset includes fully-sampled $k$-space raw data with 973 scans (34,742 slices) for training and 199 scans (7,135 slices) for validation. Additionally, a test set with 118 under-sampled $k$-space scans (4,092 slices) is provided for participants to upload their reconstructions to the public leaderboard for performance comparison. The data were acquired using a 15-channel knee coil and a Cartesian 2D TSE protocol, with following sequence parameters: echo train length 4, matrix size 320 $\times$ 320, in-plane resolution 0.5 mm $\times$ 0.5 mm, slice thickness 3 mm, no gap between slices. The dataset was acquired with proton density weighting with half having fat saturation (PDFS) and the other half having no fat suppression (PD). We performed online testing with acceleration factors of 4 and 8. Examples of the provided masks are shown in Fig.~\ref{fig:sampling_mask}~\subref{fig:sampling_mask_e} and Fig.~\ref{fig:sampling_mask}~\subref{fig:sampling_mask_f}.
	
	\subsection{Implementation Details}
	CAMP-Net is independently trained from scratch for each dataset. The number of iterations ($T$) is set to 8 for the Calgary-Campinas and SKM-TEA datasets, and 12 for the fastMRI dataset. The IEM module adopts a 4-stage U-Net architecture with initial feature dimensions of 64, 16, and 32 for the Calgary-Campinas, SKM-TEA, and fastMRI datasets, respectively. The feature dimensions of both the KRM and CCM modules are specifically matched to the number of coils in each dataset. The SEM module utilizes the network configuration from the work \cite{zbontar2018fastmri}. For 3D acquisition datasets, 4 and 2 adjacent slices are used for the Calgary-Campinas and SKM-TEA datasets, respectively. However, the adjacent slice feature is disabled for the fastMRI dataset due to its 2D acquisition nature.
	
	Models are trained using the Adam optimizer with parameters $\beta_1 = 0.9$ and $\beta_2 = 0.999$. The models are trained for 30, 20, and 50 epochs on the Calgary-Campinas, SKM-TEA, and fastMRI datasets, respectively. The initial learning rate is $3e^{-4}$ and reduced by a factor of 10 after 20, 16, and 40 epochs for the respective datasets. Implementation is done using PyTorch Lightning framework with a batch size of 1 and 4 NVIDIA RTX A6000 GPUs for training. The code is available online\footnote{\url{https://github.com/lpzhang/CAMP-Net}}.
	
	\subsection{Results on Calgary-Campinas Dataset}
	\begin{table}[!t]
		\caption{Results [mean (standard deviation)] on Calgary-Campinas dataset at 5X and 10X acceleration factors.}
		\label{tbl:calgary-campinas-5x10x}
		\centering
		\resizebox{\columnwidth}{!}{
			\begin{tabular}{lcccc}
				\toprule
				& \multicolumn{2}{c}{5X} & \multicolumn{2}{c}{10X} \\
				\cmidrule(r){2-3} \cmidrule(r){4-5}
				Model & PSNR $\uparrow$ & SSIM $\uparrow$ & PSNR $\uparrow$ & SSIM $\uparrow$ \\
				\midrule
				U-Net & 29.77 (0.68) & 0.8769 (0.0095) & 27.86 (0.52) & 0.8283 (0.0141) \\
				MultiDomainNet & 28.70 (0.63) & 0.8682 (0.0100) & 26.87 (0.64) & 0.8160 (0.0155) \\
				KIKI-Net & 29.66 (0.73) & 0.8923 (0.0094) & 27.63 (0.71) & 0.8393 (0.0145) \\
				XPDNet & 32.63 (0.42) & 0.9115 (0.0071) & 29.99 (0.55) & 0.8607 (0.0114) \\
				Joint-ICNet & 32.52 (0.63) & 0.9118 (0.0093) & 29.79 (0.61) & 0.8638 (0.0148) \\
				E2EVarNet & 33.69 (0.49) & 0.9210 (0.0068) & 30.24 (0.53) & 0.8689 (0.0114) \\
				RIM & 35.24 (0.45) & 0.9348 (0.0061) & 32.10 (0.46) & 0.8969 (0.0087) \\
				IterDualNet & 35.26 (0.39) & 0.9375 (0.0059) & 32.16 (0.40) & 0.9027 (0.0083) \\
				LPDNet & 35.71 (0.43) & 0.9388 (0.0061) & 32.51 (0.46) & 0.9052 (0.0085) \\
				ConjGradNet & 35.73 (0.42) & 0.9390 (0.0058) & 32.68 (0.42) & 0.9074 (0.0080) \\
				RecurrentVarNet & 36.35 (0.39) & 0.9446 (0.0054) & 33.33 (0.40) & 0.9159 (0.0068) \\
				CAMP-Net & \textbf{36.74 (0.50)} & \textbf{0.9476 (0.0058)} & \textbf{34.08 (0.51)} & \textbf{0.9265 (0.0070)} \\
				\bottomrule
			\end{tabular}
		}
	\end{table}
	
	\begin{figure}[!t]
		\centering{
			\includegraphics[width=0.96\linewidth]{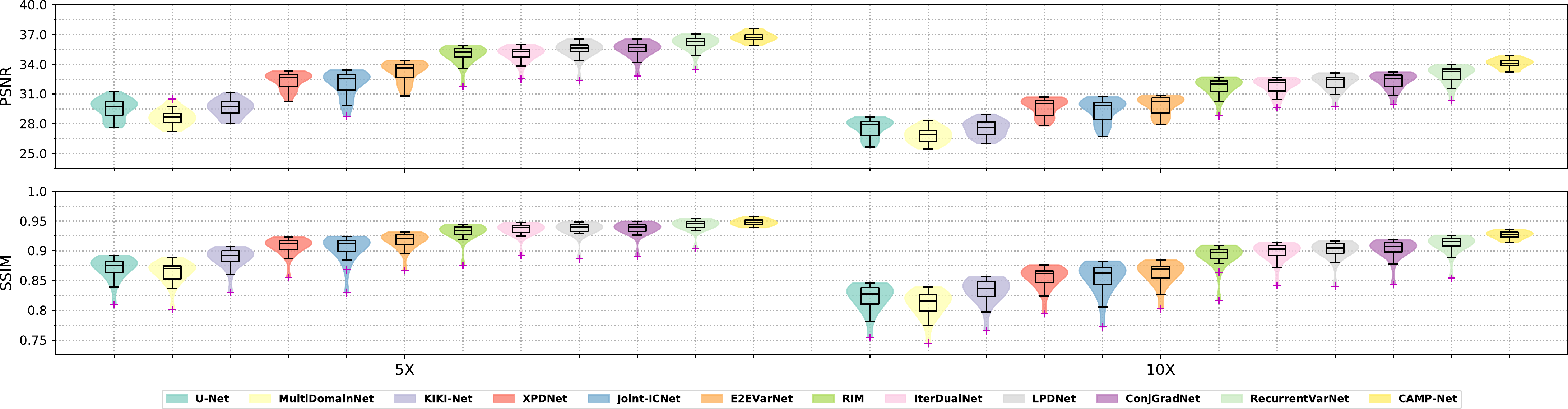}
		}
		\caption{
			Performance of reconstruction models (colored legend) for the Calgary-Campinas dataset at 5X (left) and 10X (right) acceleration.
		}
		\label{fig:calgary-campinas-violin-5x10x}
	\end{figure}
	
	\begin{figure*}[!t]
		\centerline{
			\includegraphics[width=0.96\linewidth]{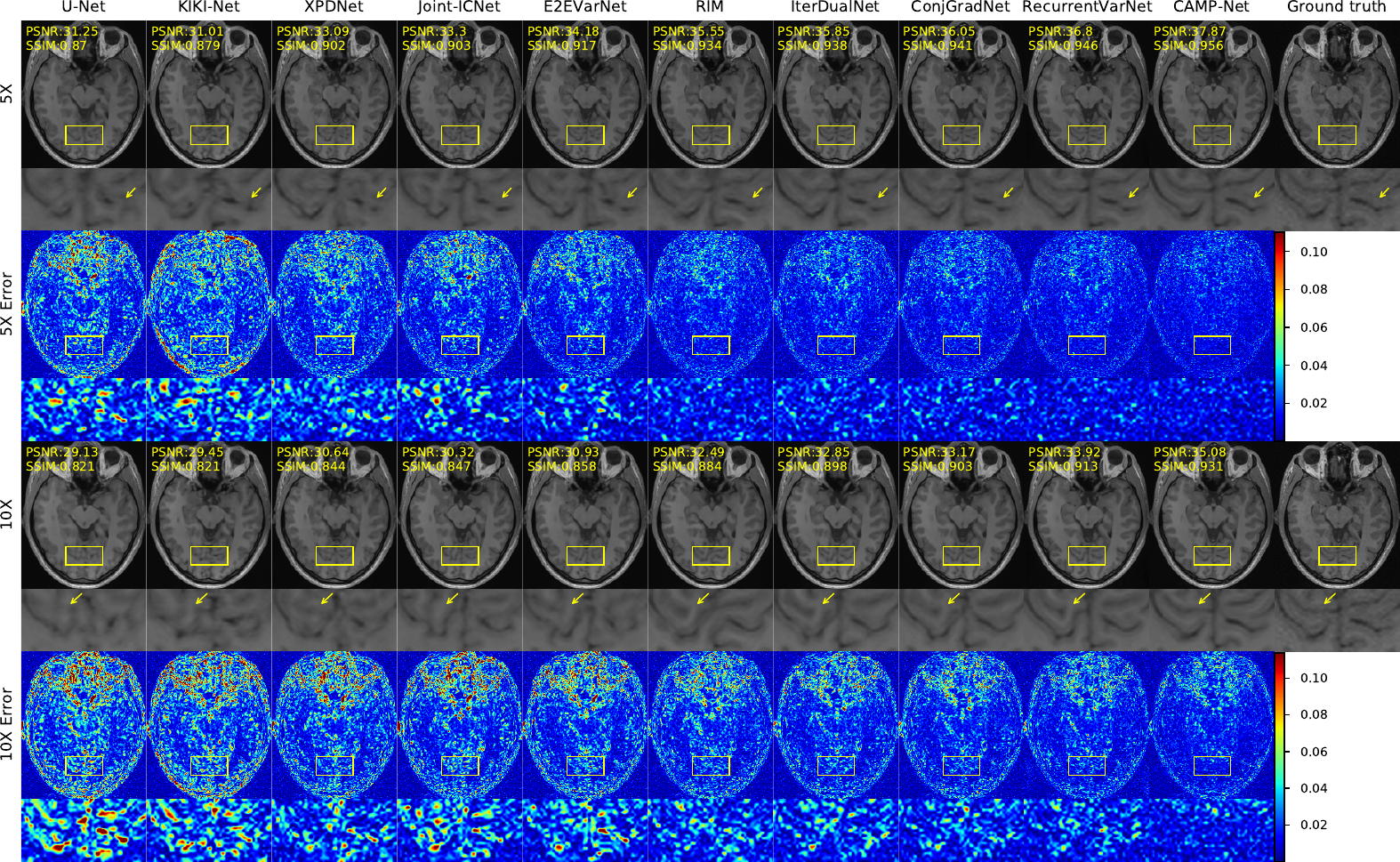}
		}
		\caption{
			Examples of brain image reconstruction and error maps on the Calgary-Campinas at 5X and 10X accelerations. Each image includes a zoomed-in view with highlighted regions of interest (yellow rectangles) for detailed analysis. The yellow arrow emphasizes restored subtle brain details.
		}
		\label{fig:calgary-campinas-5x10x}
	\end{figure*}

	CAMP-Net was compared with various state-of-the-art (SOTA) methods on the Calgary-Campinas dataset, including U-Net \cite{zbontar2018fastmri}, MultiDomainNet \cite{muckley2021results}, KIKI-Net \cite{eo2018kiki}, XPDNet \cite{ramzi2020xpdnet}, Joint-ICNet \cite{jun2021joint}, E2EVarNet \cite{sriram2020end}, RIM \cite{lonning2019recurrent}, LPDNet \cite{adler2018learned}, IterDualNet, ConjGradNet, and RecurrentVarNet \cite{yiasemis2022recurrent}. These SOTA methods were implemented using the Deep Image Reconstruction Toolkit (DIRECT) \cite{yiasemis2022direct} with released pre-trained models. Detailed configurations can be found online\footnote{\text{https://github.com/NKI-AI/direct}}.
	
	The quantitative evaluation results are shown in Table~\ref{tbl:calgary-campinas-5x10x}. CAMP-Net outperforms all methods in both PSNR and SSIM at both accelerations, with greater improvements observed at 10X acceleration. Compared to the best-published method RecurrentVarNet \cite{yiasemis2022recurrent}, CAMP-Net improves PSNR by 0.39 dB and 0.75 dB at 5X and 10X accelerations, respectively, while also consistently improving SSIM. CAMP-Net significantly outperforms E2EVarNet \cite{sriram2020end}, which uses only the image prior for reconstruction. This demonstrates the effectiveness of our multi-prior collaborative learning approach. The violin plots in Fig.~\ref{fig:calgary-campinas-violin-5x10x} illustrate the highlighted outliers and the discrepancies of the compared methods at both 5X and 10X acceleration factors. Note CAMP-Net has the highest median and no identifiable outliers.
	
	Fig.~\ref{fig:calgary-campinas-5x10x} shows reconstructed images and the associated error maps compared to the ground truth at 5X and 10X acceleration, respectively. The even rows are zoomed-in images of the regions indicated by  the yellow boxes. Note CAMP-Net achieved notable improvements in image reconstruction with the smallest errors compared to the other SOTA methods at both accelerations. The high-frequency details and fine-scale textures in the occipitial lobe (indicated by yellow arrows) are better visualized with CAMP-Net reconstruction, demonstrating the effectiveness of the proposed multi-prior collaborative learning strategy. Note that the CAMP-Net reconstruction even exhibits higher SNR levels compared to the reference image.
	
	\subsection{Results on SKM-TEA Dataset}
	\begin{table}[!t]
		\caption{
			Results [mean (standard deviation)] on the SKM-TEA dataset for echoes E1 and E2 at acceleration factors of 6X and 8X.	
		}
		\label{tbl:skmtea-e1e2-6x8x}
		\centering
		\resizebox{\columnwidth}{!}{
			\begin{tabular}{llcccc}
				\toprule
				& & \multicolumn{2}{c}{PSNR $\uparrow$} & \multicolumn{2}{c}{SSIM $\uparrow$} \\
				\cmidrule(r){3-4} \cmidrule(r){5-6}
				Acc. & Model & E1 & E2 & E1 & E2 \\
				\midrule
				\multirow{6}{*}{6X}
				& U-Net (E1/E2) & 31.54 (1.38) & 33.68 (1.02) & 0.7652 (0.0273) & 0.7254 (0.0319) \\
				& U-Net (E1+E2) & 31.07 (1.38) & 33.16 (1.05) & 0.7687 (0.0244) & 0.7345 (0.0299) \\
				& U-Net (E1$\oplus$E2) & 31.10 (1.63) & 33.46 (1.02) & 0.7612 (0.0259) & 0.7296 (0.0341) \\
				& Unrolled (E1/E2) & 35.03 (1.08) & 34.46 (1.09) & 0.8296 (0.0242) & 0.7607 (0.0309) \\
				& Unrolled (E1+E2) & 35.02 (1.07) & 34.48 (1.09) & 0.8384 (0.0222) & 0.7612 (0.0298) \\
				& Unrolled (E1$\oplus$E2) & 35.01 (1.08) & 34.16 (1.08) & 0.8333 (0.0230) & 0.7563 (0.0298) \\
				& CAMP-Net (E1/E2) & \textbf{36.30 (1.16)} & 35.85 (1.15) & \textbf{0.9071 (0.0138)} & 0.8497 (0.0282) \\
				& CAMP-Net (E1+E2) & 36.25 (1.14) & 35.87 (1.16) & 0.9063 (0.0135) & 0.8502 (0.0282) \\
				& CAMP-Net (E1$\oplus$E2) & 36.25 (1.21) & \textbf{35.92 (1.19)} & 0.9062 (0.0140) & \textbf{0.8514 (0.0286)} \\
				\midrule
				\multirow{6}{*}{8X}
				& U-Net (E1/E2) & 30.61 (1.55) & 32.91 (1.02) & 0.7266 (0.0295) & 0.6736 (0.0352) \\
				& U-Net (E1+E2) & 30.78 (1.24) & 32.51 (1.00) & 0.7219 (0.0300) & 0.6835 (0.0352) \\
				& U-Net (E1$\oplus$E2) & 30.79 (1.23) & 32.74 (1.03) & 0.7212 (0.0293) & 0.6787 (0.0386) \\
				& Unrolled (E1/E2) & 33.83 (1.07) & 33.73 (1.06) & 0.7927 (0.0274) & 0.7267 (0.0334) \\
				& Unrolled (E1+E2) & 33.82 (1.07) & 33.62 (1.07) & 0.7969 (0.0268) & 0.7064 (0.0346) \\
				& Unrolled (E1$\oplus$E2) & 33.90 (1.08) & 33.88 (1.07) & 0.7965 (0.0270) & 0.7262 (0.0327) \\
				& CAMP-Net (E1/E2) & \textbf{35.28 (1.12)} & 35.27 (1.12) & \textbf{0.8907 (0.0153)} & 0.8315 (0.0301) \\
				& CAMP-Net (E1+E2) & 35.22 (1.11) & 35.29 (1.12) & 0.8896 (0.0151) & 0.8319 (0.0301) \\
				& CAMP-Net (E1$\oplus$E2) & 35.27 (1.17) & \textbf{35.37 (1.15)} & 0.8901 (0.0156) & \textbf{0.8344 (0.0306)} \\
				\bottomrule
			\end{tabular}
		}
	\end{table}
	
	\begin{figure}[!t]
		\centering{
			\includegraphics[width=0.96\linewidth]{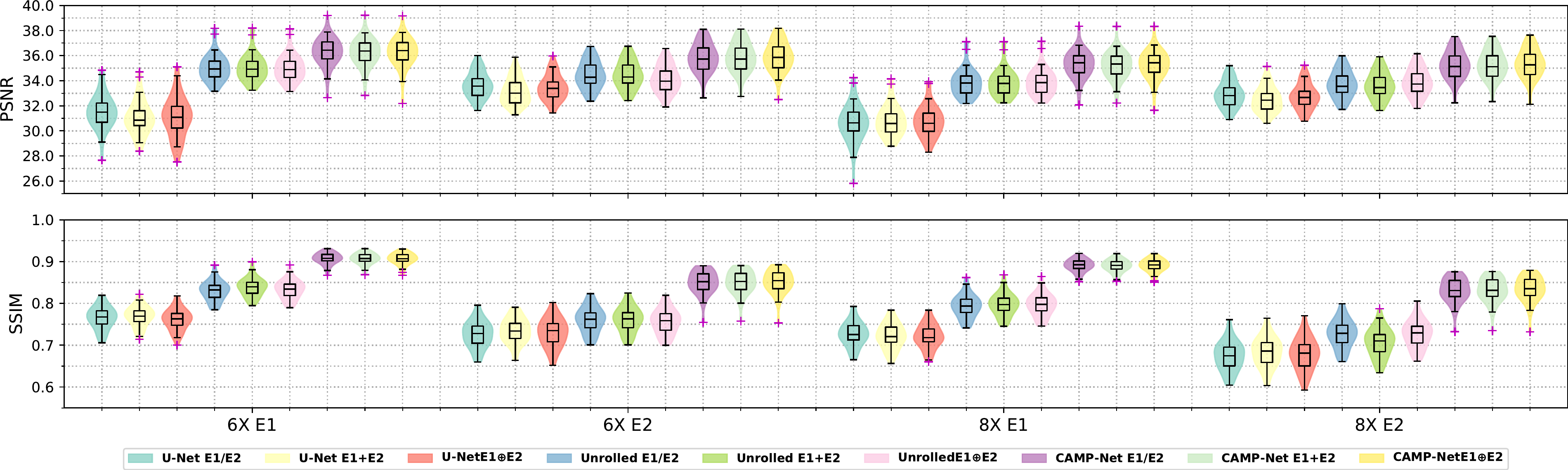}
		}
		\caption{
			Performance of reconstruction models (colored legend) on the SKM-TEA dataset for echoes E1 and E2 at 6X and 8X acceleration.
		}
		\label{fig:skmtea-recon-violin-6x8x}
	\end{figure}
	
	\begin{figure*}[!t]
		\centering
		\includegraphics[width=0.96\linewidth]{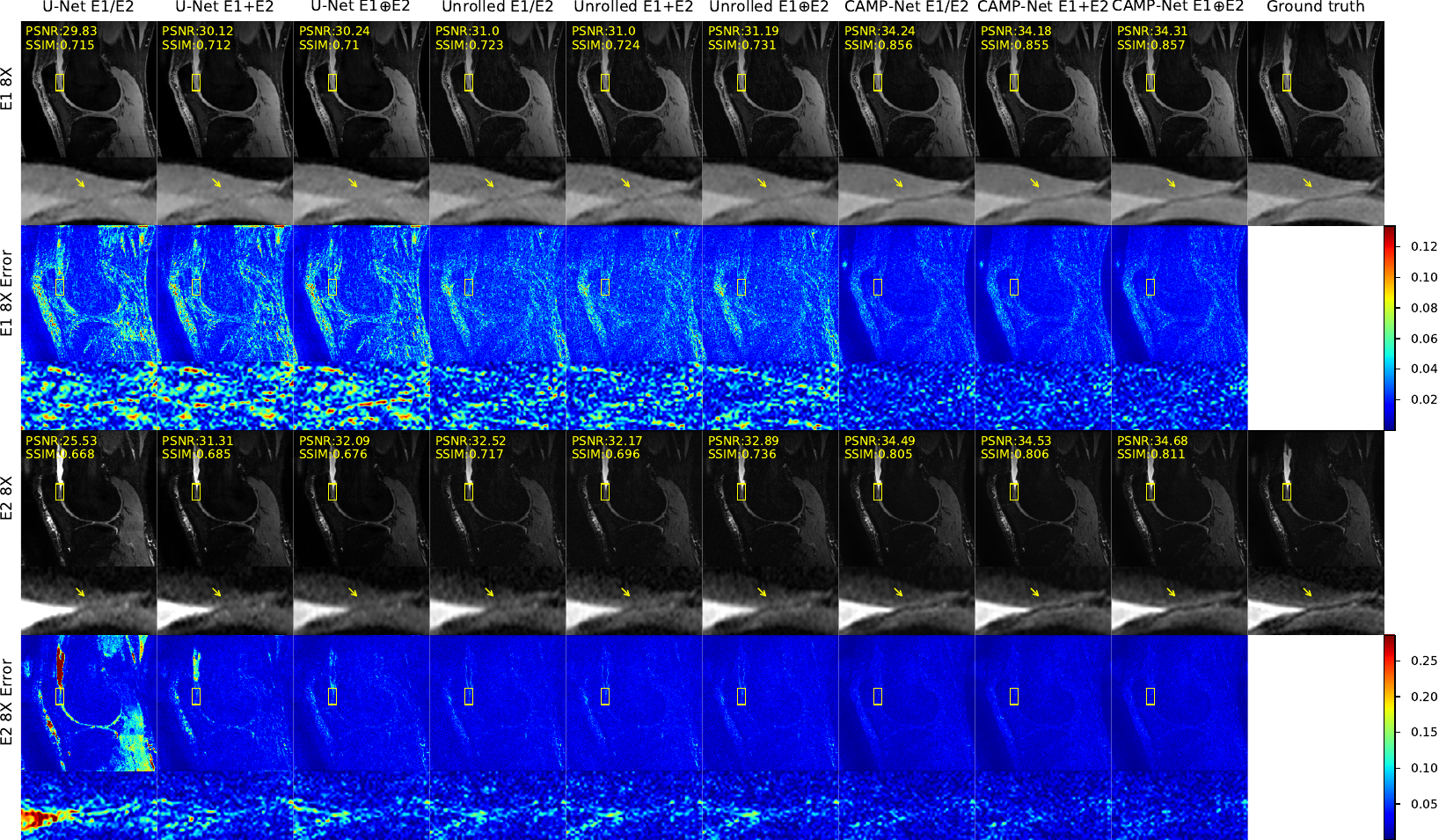}
		\caption{
			Examples of qDESS E1/E2 knee image reconstruction and error maps for the SKM-TEA at an acceleration factor of 8X. Each reconstructed image includes a zoomed-in view of regions within the yellow rectangles. Yellow arrows indicate areas with subtle structures.
		}
		\label{fig:skmtea-e1e2-8x}
	\end{figure*}
	
	We evaluated CAMP-Net's generalization capability on the SKM-TEA dataset by comparing it with baselines from the raw data reconstruction benchmark track \cite{desai2022skm}, including U-Net \cite{zbontar2018fastmri} and unrolled network \cite{sandino2020compressed}.
	Following the settings in the SKM-TEA dataset, each type of model underwent training using three configurations to reconstruct two 3D qDESS images (E1 and E2 echoes): (1) separate models for each echo (E1/E2), (2) a single model for both echoes with each echo treated as a unique training example (E1+E2), and (3) a single model for both echoes with echoes treated as multiple channels of a single example (E1$\oplus$E2). Besides PSNR and SSIM, we evaluated the reconstructed images for estimating the clinically relevant $T_2$ relaxation time, which showed higher concordance with radiologist assessments \cite{desai2022skm}.
	The model-generated qDESS images and ground-truth tissue segmentation are used to estimate tissue-wise $T_2$ maps, including patellar cartilage (PC), femoral cartilage (FC), tibial cartilage (TC), and meniscus (Men).
	
	Table \ref{tbl:skmtea-e1e2-6x8x} presents the quantitative evaluation results. Our models achieve significantly higher PSNR and SSIM compared to the benchmarks for both echoes and acceleration factors across the three data configurations. Compared to other methods using the same architecture, CAMP-Net is less sensitive to data configurations and echo types, resulting in a more consistent performance for both echoes. Fig.~\ref{fig:skmtea-recon-violin-6x8x} illustrates the improvement of CAMP-Net over the baselines for both echoes and acceleration factors. Fig.~\ref{fig:skmtea-e1e2-8x} shows reconstructed images and error maps for qDESS E1 and E2 at an acceleration factor of 8X. Notably, reconstructing the cartilage surface, which often consists of only a few pixels, is a challenging task. The benchmarks show blurry reconstruction in the surface of the PC and FC for both echoes across all data configurations, leading to a narrowed articular space, as indicated by the yellow arrows. In contrast, CAMP-Net successfully reconstructs these subtle structural details, demonstrating its potential in restoring high-frequency information.
	
	\begin{table*}
		\caption{
			Quantitative evaluation results [mean (standard deviation)] of the SKM-TEA reconstruction with respect to absolute $T_2$ error (in milliseconds) for articular cartilage and the meniscus localized with ground truth segmentation.
		}
		\label{tbl:skmtea-t2-6x8x}
		\centering
		\resizebox{\textwidth}{!}{
			\begin{tabular}{l cccc cccc}
				\toprule
				& \multicolumn{4}{c}{6X} & \multicolumn{4}{c}{8X} \\
				\cmidrule(r){2-5} \cmidrule(r){6-9}
				Model & PC & FC & TC & Men &PC & FC & TC & Men  \\
				\midrule
				U-Net (E1/E2) & 2.189 (1.677) & 1.077 (0.939) & 1.615 (0.948) & 2.698 (1.350) & 3.481 (1.741) & 2.711 (1.375) & 3.207 (1.242) & 3.763 (1.101) \\
				U-Net (E1+E2) & 2.831 (1.949) & 2.460 (1.877) &  1.462 (0.917) & 2.009 (1.418) & 2.659 (2.061) & 3.037 (2.027) & 1.486 (1.157) & 2.387 (1.308) \\
				U-Net (E1$\oplus$E2) & 1.769 (1.503) & 1.115 (0.782) & 1.538 (1.029) & 1.814 (0.966) & 1.291 (1.088) &  1.262 (0.913) & 2.085 (1.131) & 2.494 (1.800) \\
				Unrolled (E1/E2) & 0.563 (0.233) &  0.765 (0.283) & 1.030 (0.419) &  2.479 (0.786) & 0.721 (0.297) & 0.899 (0.336) & 1.259 (0.486) & 2.779 (0.868) \\
				Unrolled (E1+E2) & 0.570 (0.233) &  0.836 (0.319) & 1.124 (0.421) & 2.519 (0.780) & 0.971 (0.419) &  0.988 (0.394) & 1.295 (0.494) & 2.862 (0.882) \\
				Unrolled (E1$\oplus$E2) & 1.685 (1.362) & 2.013 (0.917) & 1.341 (0.555) & 1.310 (0.819) & 0.588 (0.292) & 0.992 (0.432) & 1.332 (0.626) & 2.729 (0.891) \\
				CAMP-Net (E1/E2) & 0.434 (0.377) & 0.445 (0.483) & 0.481 (0.472) & 0.486 (0.394) & 0.420 (0.411) & 0.480 (0.557) & 0.551 (0.645) & 0.736 (0.658) \\
				CAMP-Net (E1+E2) & \textbf{0.259 (0.266)} & \textbf{0.361 (0.455)} & \textbf{0.432 (0.463)} & 0.536 (0.299) & \textbf{0.290 (0.321)} & \textbf{0.423 (0.459)} & \textbf{0.488 (0.480)} & 0.759 (0.463) \\
				CAMP-Net (E1$\oplus$E2) & 0.311 (0.349) & 0.474 (0.584) & 0.449 (0.532) & \textbf{0.470 (0.374)} & 0.363 (0.369) & 0.566 (0.724) & 0.517 (0.655) & \textbf{0.568 (0.404)} \\
				\bottomrule
			\end{tabular}
		}
	\end{table*}
	
	\begin{figure}
		\centering{
			\includegraphics[width=0.96\linewidth]{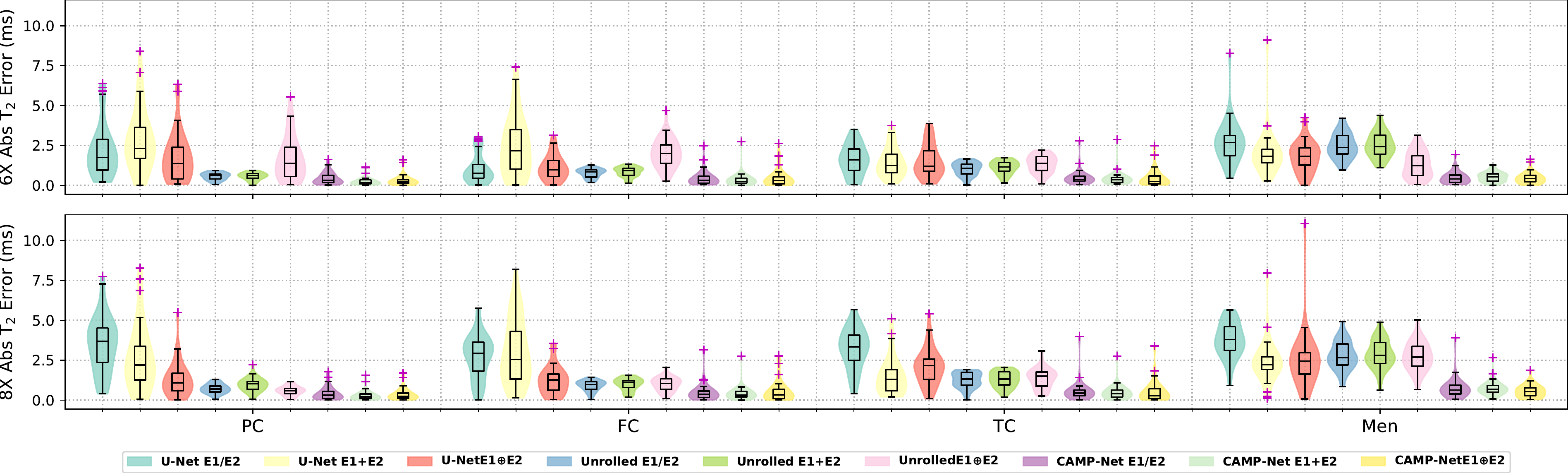}
		}
		\caption{
			Absolute $T_2$ estimation error (in milliseconds) for different reconstruction models (colored legend) across PC, FC, TC, and Men tissues.
		}
		\label{fig:skmtea-t2-violin-6x8x}
	\end{figure}
	
	\begin{figure*}
		\centerline{
			\includegraphics[width=0.96\linewidth]{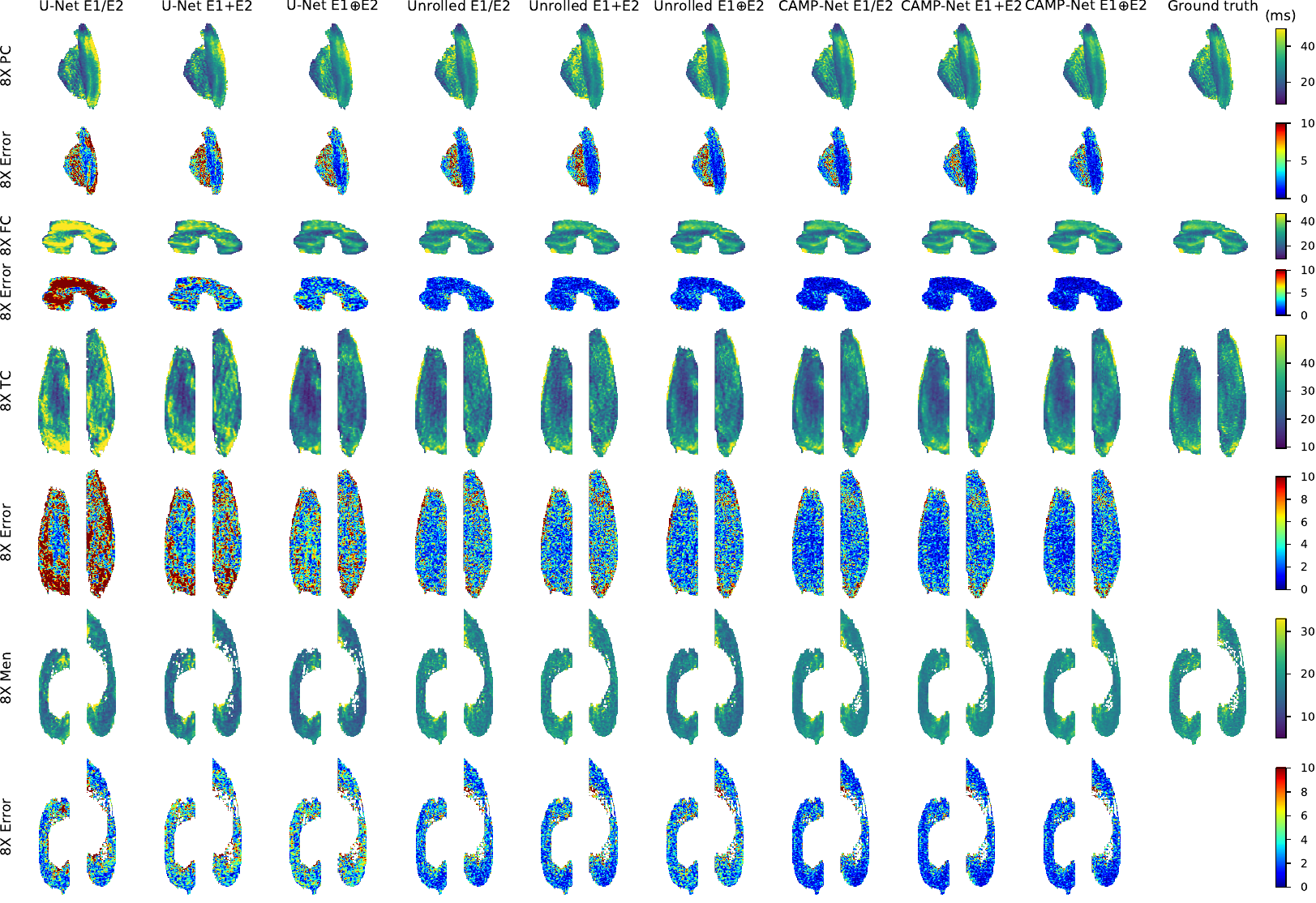}
		}
		\caption{
			Examples of $T_2$ estimation and the corresponding error maps (in milliseconds) for PC, FC, TC, and Men tissues at an acceleration factor of 8X.
		}
		\label{fig:skmtea-t2-6x8x}
	\end{figure*}
	
	Table \ref{tbl:skmtea-t2-6x8x} presents the evaluation results of absolute $T_2$ error (in milliseconds).
	CAMP-Net achieves superior $T_2$ quantification results in all tissues at both accelerations compared to other methods, regardless of data configurations. Furthermore, compared to the baseline, the performance of CAMP-Net evaluated by the standard metrics and $T_2$ quantification is more consistent. We also employed violin plots as a visual tool to effectively summarize and compare the distributions of the results generated by models with different input settings for PC, FC, TC, and Men tissues, as shown in Fig.~\ref{fig:skmtea-t2-violin-6x8x}. We observed that the meniscus is the most challenging tissue to achieve accurate $T_2$ map estimation compared to PC, FC, and TC when utilizing the reconstructed images from the baselines. Nevertheless, the reconstruction from CAMP-Net consistently provides better $T_2$ quantification for all tissues compared to the other methods. Fig.~\ref{fig:skmtea-t2-6x8x} shows examples of $T_2$ estimation and the corresponding error maps for different tissues at an acceleration factor of 8X. It is evident that CAMP-Net yields $T_2$ maps that are more closely aligned with the ground truth compared to the other methods.
	
	\subsection{Results on fastMRI Knee Dataset}
	\begin{figure*}
		\centerline{
			\includegraphics[width=0.96\linewidth]{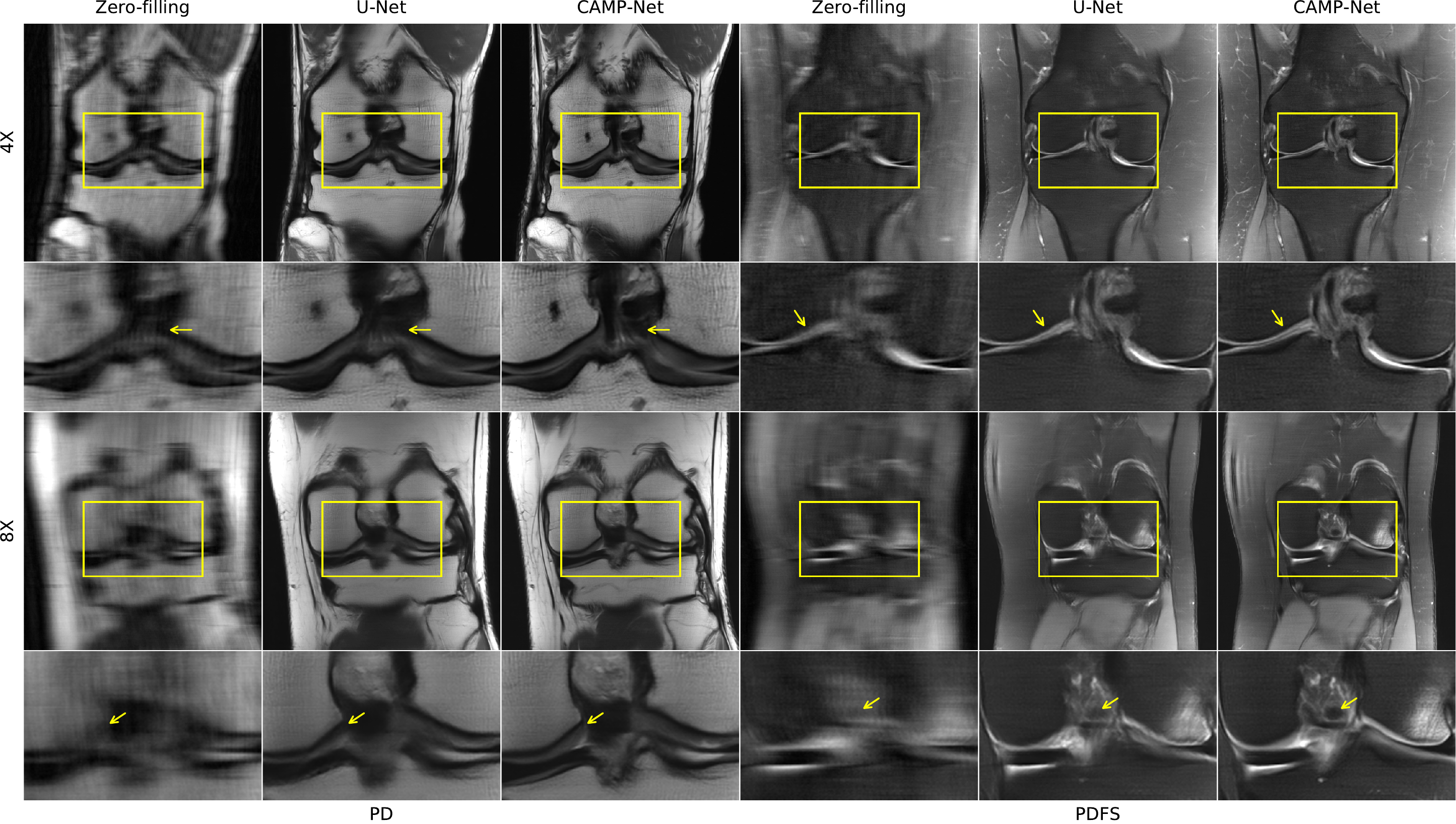}
		}
		\caption{
			Examples of reconstruction on the unseen test data from the fastMRI dataset for PD and PDFS at acceleration factors of 4X and 8X.
		}
		\label{fig:fastmri-4x8x}
	\end{figure*}
	
	\begin{table*}
		\caption{
			Public Leaderboard Results for 4X and 8X acceleration (Acc.) on the fastMRI Multi-Coil Knee dataset.
		}
		\label{tbl:public-leaderboard-fastmri-knee-mc}
		\centering
			\begin{tabular}{l cccccc cccccc}
				\toprule
				& \multicolumn{6}{c}{4X} & \multicolumn{6}{c}{8X} \\
				\cmidrule(r){2-7} \cmidrule(r){8-13}
				Model & \multicolumn{3}{c}{SSIM $\uparrow$} & \multicolumn{3}{c}{PSNR $\uparrow$} & \multicolumn{3}{c}{SSIM $\uparrow$} & \multicolumn{3}{c}{PSNR $\uparrow$} \\
				\cmidrule(r){2-4} \cmidrule(r){5-7} \cmidrule(r){8-10} \cmidrule(r){11-13}
				& ALL & PD & PDFS & ALL & PD & PDFS & ALL & PD & PDFS & ALL & PD & PDFS \\
				\midrule
				\textbf{CAMP-Net} & \textbf{0.9320} & \textbf{0.9642} & \textbf{0.8974} & \textbf{40.3} & \textbf{42.1} & \textbf{38.3} & \textbf{0.8949} & \textbf{0.9430} & \textbf{0.8499} & \textbf{37.4} & \textbf{39.2} & \textbf{35.8} \\
				HUMUS-Net & 0.9238 & 0.9556 & 0.8898 & 38.2 & 40.0 & 36.3 & 0.8945 & 0.9419 & 0.8501 & 37.3 & 39.0 & 35.7 \\
				E2EVarNet & 0.9302 & 0.9619 & 0.8962 & 39.9 & 41.6 & 38.1 & 0.8920 & 0.9393 & 0.8477 & 37.1 & 38.6 & 35.6 \\
				i-RIM & 0.9278 & 0.9592 & 0.8942 & 39.6 & 41.2 & 38.0 & 0.8875 & 0.9338 & 0.8442 & 36.7 & 38.0 & 35.4 \\
				$\Sigma$-Net & 0.9282 & 0.9611 & 0.8929 & 39.8 & 41.6 & 38.0 & 0.8877 & 0.9354 & 0.8431 & 36.7 & 38.2 & 35.4 \\
				XPDNet & 0.9287 & 0.9623 & 0.8926 & 40.2 & 42.0 & 38.2 & 0.8893 & 0.9383 & 0.8435 & 37.2 & 38.8 & 35.7 \\
				U-Net & 0.9103 & 0.9377 & 0.8808 & 37.6 & 38.2 & 37.0 & 0.8640 & 0.9054 & 0.8253 & 34.7 & 35.3 & 34.2 \\		
				\bottomrule
			\end{tabular}
	\end{table*}
	
	We further evaluated our model on the fastMRI Knee dataset and compared it with the published SOTA methods listed on the public leaderboard, including HUMUS-Net \cite{fabian2022humus}, E2EVarNet \cite{sriram2020end}, i-RIM \cite{putzky2019rim}, $\Sigma$-Net \cite{hammernik2021systematic}, XPDNet \cite{ramzi2020xpdnet}, and U-Net \cite{zbontar2018fastmri}. The online evaluation results are presented in Table \ref{tbl:public-leaderboard-fastmri-knee-mc}, which can be accessed via the archived historic leaderboards\footnote{\url{https://web.archive.org/web/20230324102125mp_/https://fastmri.org/leaderboards}}. CAMP-Net achieves the best performance in both accelerations. Fig.~\ref{fig:fastmri-4x8x} shows representative reconstruction examples of PD and PDFS images from the unseen test data at 4X and 8X acceleration factors.
	CAMP-Net provides improved visualization of structural details in reconstructed images compared to the other methods. The boundaries of the articular cartilage and cruciate ligament (indicated by yellow arrows) are depicted more accurately with CAMP-Net reconstruction at both acceleration factors for PD and PDFS acquisitions. These results demonstrate the potential of generalizability of CAMP-Net in restoring and preserving fine-scale textures and high-frequency structural details.
	
	\subsection{Ablation Study}
	\begin{table*}[!t]
		\caption{
			Ablation study results on the Calgary-Campinas dataset evaluating the performance of different settings.
		}
		\label{tbl:calgary-campinas-ablation-5x10x}
		\centering
		\resizebox{\textwidth}{!}{
			\begin{tabular}{l|cccccccccc}
				\toprule
				\multirow{2}{*}{Models} & \multirow{2}{*}{\text{IEM}} & \multirow{2}{*}{\text{KRM}} & \multirow{2}{*}{\text{CCM}} & \multirow{2}{*}{\text{FFM}} & \multirow{2}{*}{\text{SDF}} & \multirow{2}{*}{\text{AS}} & \multicolumn{2}{c}{5X} & \multicolumn{2}{c}{10X} \\
				\cmidrule(r){8-9} \cmidrule(r){10-11}
				& & & & & & & PSNR $\uparrow$ & SSIM $\uparrow$ & PSNR $\uparrow$ & SSIM $\uparrow$ \\
				\midrule
				CAMP-Net-4AS & $\surd$ &$\surd$ & $\surd$ & $\surd$ & $\surd$ & 4 & \textbf{36.74 (0.50)} & \textbf{0.9476 (0.0058)} & \textbf{34.08 (0.51)} & \textbf{0.9265 (0.0070)} \\
				CAMP-Net-wo-SDF & $\surd$ &$\surd$ & $\surd$ & $\surd$ & & 4 & 36.10 (0.79) & 0.9433 (0.0075) & 33.33 (0.72) & 0.9191 (0.0093) \\
				CAMP-Net-wo-FFM & $\surd$ & $\surd$ & $\surd$ & & $\surd$ & 4 & 35.51 (1.20) & 0.9387 (0.0104) & 32.70 (1.00) & 0.9109 (0.0130) \\
				CAMP-Net-wo-CCM & $\surd$ & $\surd$ & & $\surd$ & $\surd$ & 4 & 35.82 (1.26) & 0.9412 (0.0107) & 33.15 (1.11) & 0.9166 (0.0136) \\
				CAMP-Net-wo-CCM-FFM & $\surd$ & $\surd$ & & & $\surd$ & 4 & 35.73 (1.20) & 0.9414 (0.0091) & 33.06 (0.96) & 0.9167 (0.0113) \\
				Baseline-4AS & $\surd$ & & & & & 4 & 32.74 (2.91) & 0.9082 (0.0399) & 30.83 (2.34) & 0.8832 (0.0413) \\
				\midrule
				CAMP-Net-2AS & $\surd$ &$\surd$ & $\surd$ & $\surd$ & $\surd$ & 2 & 36.68 (0.44) & 0.9473 (0.0054) & 33.86 (0.45) & 0.9241 (0.0066) \\
				CAMP-Net-0AS & $\surd$ &$\surd$ & $\surd$ & $\surd$ & $\surd$ & 0 & 33.75 (2.15) & 0.9171 (0.0277) & 30.84 (1.79) & 0.8793 (0.0328) \\
				Baseline-2AS & $\surd$ & & & & & 2 & 33.46 (2.07) & 0.9190 (0.0253) & 31.27 (1.55) & 0.8929 (0.0238) \\
				Baseline & $\surd$ & & & & & 0 & 32.37 (2.41) & 0.9021 (0.0352) & 30.08 (1.83) & 0.8665 (0.0362) \\
				\bottomrule
			\end{tabular}
		}
	\end{table*}
	
	Ablation analysis was conducted on the Calgary-Campinas dataset to evaluate the effectiveness of the core modules and the utilization of adjacent slices in CAMP-Net. The study started with CAMP-Net-4AS, which incorporates four adjacent slices, and involved the systematic removal of the KRM, CCM, FFM, and SDF modules, as well as a reduction in the number of adjacent slices. Note that the effectiveness of the SEM module has been previously demonstrated in \cite{sriram2020end}. We performed a paired t-test with a significance level of $0.05$.
	
	The results of the ablation study at 5X and 10X acceleration factors are presented in Table~\ref{tbl:calgary-campinas-ablation-5x10x}, and the violin plots are depicted in Fig.\ref{fig:calgary-campinas-ablation-violin-5x10x}. Additionally, Fig.\ref{fig:calgary-campinas-ablation-5x10x} displays representative reconstructions at a 10X acceleration factor. Note that the baseline model is the 3D convolution version of E2EVarNet \cite{sriram2020end}, which solely employs the IEM module for learning image prior knowledge in MRI reconstruction. Among the compared models, CAMP-Net-4AS demonstrates the best performance in terms of both PSNR and SSIM at both acceleration factors. Omitting any of the modules results in a degradation of reconstruction quality ($p\text{-value} \ll 0.05$), highlighting the significant benefits of these modules and the utilization of multiple adjacent slice information within the CAMP-Net for MRI reconstruction.
	
	\subsubsection{Effectiveness of KRM and CCM}
	\begin{figure}[!t]
		\centering{
			\includegraphics[width=0.96\linewidth]{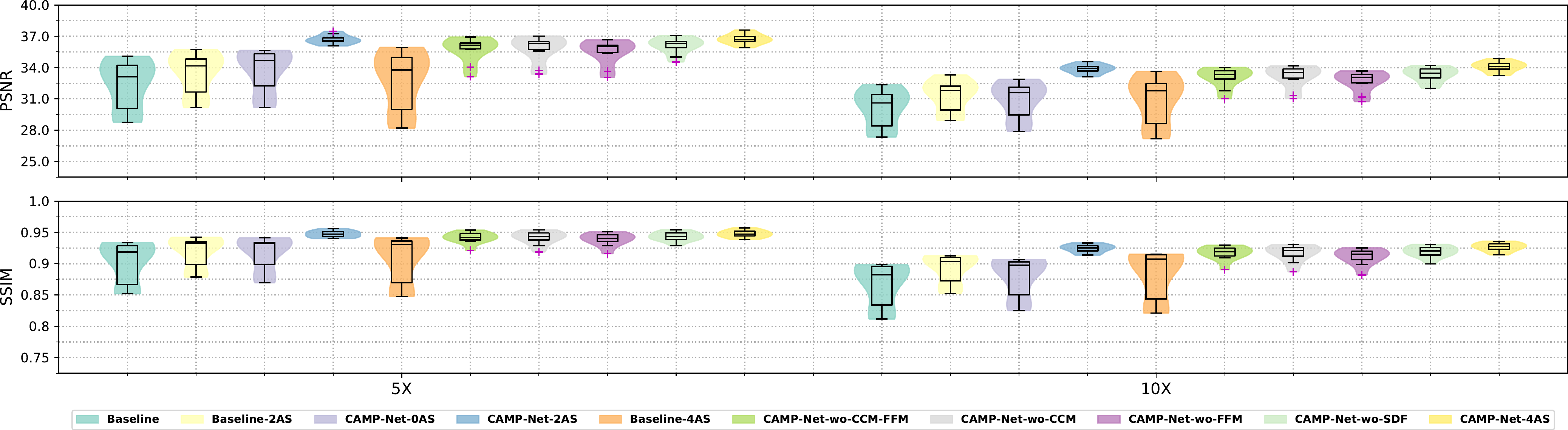}
		}
		\caption{
			Performance of the ablation study on the Calgary-Campinas dataset at 5X (left) and 10X (right) accelerations.
		}
		\label{fig:calgary-campinas-ablation-violin-5x10x}
	\end{figure}
	
	\begin{figure*}[!t]
		\centerline{
			\includegraphics[width=0.96\linewidth]{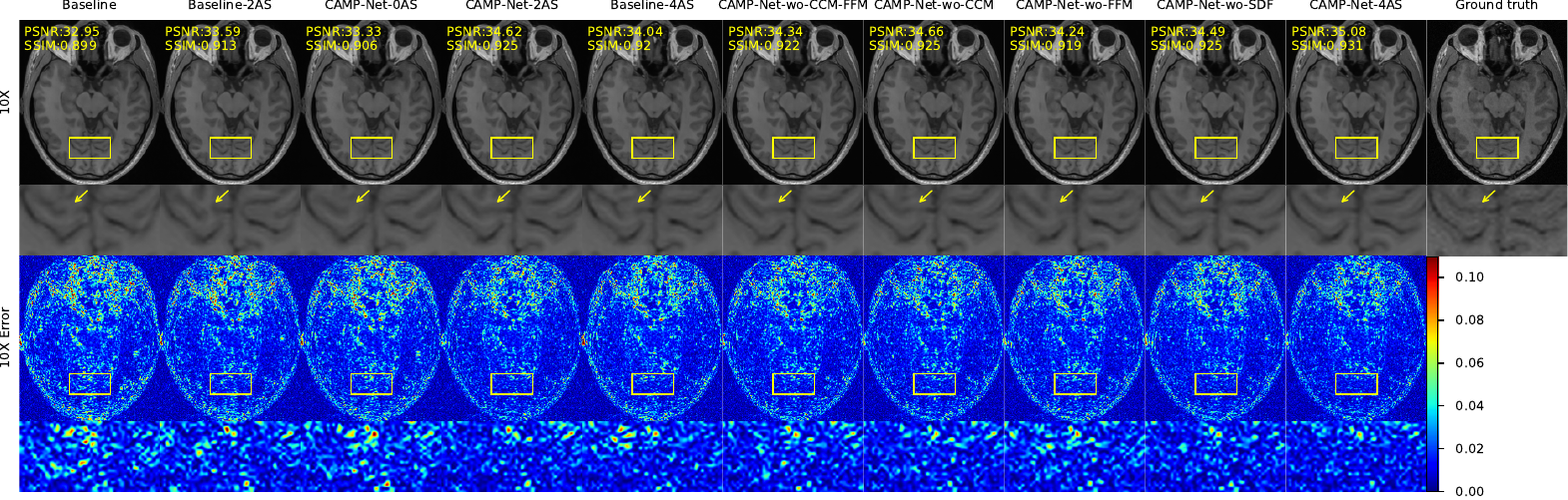}
		}
		\caption{
			Examples of brain image reconstruction and error maps of the ablation study on the Calgary-Campinas dataset at 10X accelerations. Each image includes a zoomed-in view with highlighted regions (yellow rectangles) for closer inspection. The yellow arrow emphasizes restored subtle brain details.
		}
		\label{fig:calgary-campinas-ablation-5x10x}
	\end{figure*}
	
	The removal of the CCM module from the best-performing model, CAMP-Net-4AS, results in a noticeable decline at both acceleration factors ($p\text{-value} \ll 0.05$). This decline is evident in the performance of the model CAMP-Net-wo-CCM, as indicated in Table~\ref{tbl:calgary-campinas-ablation-5x10x}. Consistent with these findings, the orange violin plot in Fig.~\ref{fig:calgary-campinas-ablation-violin-5x10x} also shows a decrease in the median and an increase in outliers for the CAMP-Net-wo-CCM model. Furthermore, the presence of blurry boundaries between the occipital sulcus and gyrus, highlighted by the yellow arrow in the seventh column of Fig.~\ref{fig:calgary-campinas-ablation-5x10x}, further supports these observations. These results emphasize the importance of incorporating calibration knowledge to guide the KRM module in capturing consistency-aware $k$-space correlations for reliable reconstruction.
	
	Similarly, when the KRM module is removed from CAMP-Net-4AS, it gives rise to a single-image prior model called Baseline-4AS, which exhibits a significant decrease in performance at both accelerations ($p\text{-value} \ll 0.05$). This decline is also evident in the violin plot shown in Fig.~\ref{fig:calgary-campinas-ablation-violin-5x10x}. Moreover, the fifth column of Fig.~\ref{fig:calgary-campinas-ablation-5x10x} highlights the presence of misleading structures generated by the degraded Baseline-4AS model, as indicated by the yellow arrow. Note that the removal of the CCM and FFM modules is a consequence of removing the KRM module itself, as these modules are associated with it. Furthermore, even in the absence of nearby slice information, CAMP-Net-0AS still outperforms all single prior models ($p\text{-value} \ll 0.05$), including Baseline-0AS, Baseline-2AS, and Baseline-4AS.
	This finding further emphasizes the importance of multiple prior knowledge for accurate reconstruction.
	
	\subsubsection{Effectiveness of FFM}
	The comparison between CAMP-Net-wo-FFM and CAMP-Net-4AS reveals notable performance drops at both acceleration factors ($p\text{-value} \ll 0.05$), as consistently shown in Table~\ref{tbl:calgary-campinas-ablation-5x10x} and the purple violin plot in Fig.~\ref{fig:calgary-campinas-ablation-violin-5x10x}. Moreover, the absence of the FFM module in CAMP-Net-wo-FFM leads to incorrect reconstruction of the occipital gyrus, as illustrated in the eighth column of Fig.~\ref{fig:calgary-campinas-ablation-5x10x}.
	Furthermore, despite the inclusion of triple priors, CAMP-Net-wo-FFM produces slightly worse reconstructions compared to the dual prior model CAMP-Net-wo-CCM ($p\text{-value} \ll 0.05$), which incorporates the FFM module.
	These results emphasize the crucial role of the FFM in integrating information from different modules by leveraging distinct properties and capturing complementary features for improved reconstruction.
	
	\subsubsection{Effectiveness of SDF}
	Table~\ref{tbl:calgary-campinas-ablation-5x10x} demonstrates that CAMP-Net-wo-SDF performs worse than CAMP-Net-4AS at both acceleration factors ($p\text{-value} \ll 0.05$), emphasizing the crucial role of the SDF module in mitigating the adverse impact of data imperfections induced by padding operations during $k$-space correlation modeling. The corresponding light green violin plot in Fig.~\ref{fig:calgary-campinas-ablation-violin-5x10x} exhibits significant deviation and outliers compared to CAMP-Net-4AS, providing further evidence of the effectiveness of the SDF module. Additionally, as shown in Fig.~\ref{fig:calgary-campinas-ablation-5x10x}, it is evident that the inclusion of the SDF module leads to sharper and clearer structures of the occipital sulcus and gyrus. This observation highlights the benefits of the SDF module in maintaining fidelity and consistency throughout the reconstruction process. By addressing the padding-induced artificial signals, the SDF module enables the model to leverage $k$-space correlations for improved reconstruction.
	
	\subsubsection{Effectiveness of AS}
	The results presented in Table~\ref{tbl:calgary-campinas-ablation-5x10x} and Fig.~\ref{fig:calgary-campinas-ablation-violin-5x10x} consistently demonstrate the improved performance of CAMP-Net when incorporating adjacent slices. Specifically, the improvement from CAMP-Net-0AS to CAMP-Net-2AS ($p\text{-value} \ll 0.05$), as depicted in the third and fourth columns of Fig.~\ref{fig:calgary-campinas-ablation-5x10x}, demonstrates the ability to preserve detailed structures by leveraging inter-slice information. This finding highlights the effectiveness of incorporating adjacent slice information into CAMP-Net to achieve improved reconstruction. Adjacent slice information also benefits the single-prior Baseline, as evidenced by the improved performance achieved by Baseline-2AS ($p\text{-value} \ll 0.05$). However, Baseline-4AS, with more adjacent slices, shows degraded performance compared to Baseline-2AS ($p\text{-value} \ll 0.05$). This is likely because the inclusion of adjacent slices also introduces irrelevant information during the reconstruction process, which can degrade the image quality when using the baseline model. In contrast, CAMP-Net shows improved robustness in utilizing adjacent slices in reconstruction. These findings demonstrate that CAMP-Net is more robust and effective in extracting useful inter-slice features for reconstruction.

	\section{DISCUSSION}
	\label{sec:discussion}
	Experiments on three publicly available MRI datasets with various acceleration factors demonstrated CAMP-Net's ability to reconstruct high-quality images with improved anatomical details and fine-scale textures at high acceleration factors.
	Several facts of the CAMP-Net contribute to its performance. Firstly, incorporating multiple learned priors in an unrolled framework provides constraints to reduce the solution space through the interaction between these priors across iterations, guiding the network to achieve progressive quality improvements in reconstruction. Secondly, the utilization of the priors learned from the calibration data in guiding the learning of consistency-aware $k$-space correlations ensures the reliable estimation of missing signals. Lastly, leveraging slice information across priors allows CAMP-Net to capture both intra- and inter-slice dependencies for enhanced image, $k$-space, and calibration prior knowledge, further improving reconstruction.
	
	This study explored how incorporating adjacent slice correlations across priors affected reconstruction quality. The results showed this approach improved reconstruction compared to single-slice methods. However, gains leveled off with more adjacent slices used in CAMP-Net. Determining the optimal number of slices is a trade off between the performance and the computational cost. The optimal number of slices may also depend on the specifics of MRI acquisitions, such as the slice thickness in 2D acquisitions and the slice resolution in 3D acquisitions. For instance, given GPU memory constraints, up to 4 adjacent slices were used for the 3D Calgary-Campinas data and 2 for the SKM-TEA data. In contrast, adjacent slices were not used for 2D fast MRI dataset to avoid potential quality degradation caused by the larger slice thickness.
	
	CAMP-Net has shown promising results in MRI reconstruction, but there are limitations and opportunities for further improvement. One limitation is that the training and testing data need to have the same coil number, which may be violated if the training and testing data are acquired using different coil arrays. Although coil compression techniques can address this issue, they may result in information loss and reduced image quality. Future work could explore integrating coil compression as a learnable layer in deep learning models to potentially overcome this limitation. Another limitation is the fixed weights of the learned priors during inference, which may cause suboptimal reconstruction due to inconsistencies between training and testing data domains. Scan-specific prior knowledge captured by the CCM module of the proposed CAMP-Net can be fine-tuned for each individual scan but involves computationally intensive inference optimization. To address this issue, future work could explore the use of domain-adaptive priors that can be adjusted to different data distributions and provide more accurate reconstructions.
	
	During the study, a 1D random under-sampling pattern was used for the 2D fastMRI knee dataset while a Poisson disc sampling pattern was used for the 3D datasets. These under-sampling patterns were selected because they are commonly adopted in the field for fair comparisons or provided with the original datasets for valid evaluation. Future work could investigate how tolerant the network is to arbitrary under-sampling patterns, which provide different properties of aliasing artifacts. Overall, such investigations have the potential to further improve the performance and applicability of CAMP-Net in MRI reconstruction, and help address challenges associated with reconstruction of under-sampled MRI data.
	
	\section{CONCLUSION}
	\label{sec:conclusion}
	In this study, we present CAMP-Ne for reconstruction of accelerated MRI. It provides a systematic framework that incorporates consistency-aware multiple prior information and leverages deep learning techniques to acquire complementary features. This enables high-quality restoration of subtle anatomical structures in highly accelerated MRI.
	Our experiments on three public datasets with various accelerations and sampling patterns validate the effectiveness of the proposed technique, demonstrating performance improvements compared to SOTA methods. Furthermore, this approach holds potential to be extend to applications in broader imaging challenges, such as dynamic MRI reconstruction, image super-resolution, denoising, and motion artifact correction.

	\bibliographystyle{IEEEtran}
	\bibliography{IEEEabrv,ref}

\end{document}